\documentclass[11pt]{elsarticle}
\usepackage{amssymb}
\usepackage{theorem}
\usepackage[british]{babel}
\usepackage[mathscr]{euscript}
\usepackage{amsmath}
\usepackage{array}
\usepackage{xspace}

\usepackage[left=2.7cm,top=3.5cm,right=2.7cm,nohead,foot=1cm]{geometry}
\usepackage{amsfonts}
\usepackage[all]{xy}
\usepackage{url}
\usepackage{ifpdf}
\ifpdf
\else
  \usepackage{graphicx}
  \DeclareGraphicsExtensions{.eps,.jpg,.png}
\fi
\usepackage{epstopdf}
\usepackage{graphicx}
\usepackage[left=2.7cm,top=3.5cm,right=2.7cm,nohead,foot=1cm]{geometry}
\graphicspath{{Figures/}}

\usepackage[hyperindex, a4paper=true]{hyperref}

\newcommand{\rmd}{{\mathrm{d}}}

\newcommand{\Real}{\mathbb{R}}
\newcommand{\Pbb}{\mathbb{P}}

\newcommand{\E}{\operatorname{\mathbb{E}}}

\def\eqd{\stackrel{d}{=}}

\usepackage{amsfonts}

\usepackage{amsmath}
\usepackage[all]{xy}
\usepackage{graphics}
\usepackage{amssymb}
\usepackage{url}
\usepackage{color}
\newenvironment{modification}{}{}
\renewenvironment{modification}{\color{red}}{}

\def\VAR{{\rm VaR}}

\theoremstyle{plain}

\newtheorem{Corollary}{Corollary}[section]

\newtheorem{Proposition}{Proposition}[section]
\newtheorem{Definition}{Definition}[section]
\newtheorem{Remark}{Remark}
\newtheorem{Example}{Example}

\begin{document}

\begin{frontmatter}
\title{On Multivariate Extensions of Value-at-Risk}

\author{Areski Cousin\footnote{Universit\'{e} de Lyon, Universit\'{e} Lyon $1$, ISFA, Laboratoire SAF, $50$ avenue Tony Garnier, $69366$ Lyon, France, Tel.: $+33  4 37 28 74 39$, areski.cousin@univ-lyon1.fr, http://www.acousin.net/.},     Elena Di Bernardino\footnote{CNAM, Paris,  Département IMATH, 292 rue Saint-Martin, Paris Cedex 03,  France.   elena.di\_bernardino@cnam.fr. http://isfaserveur.univ-lyon1.fr/\~elena.dibernardino/.}}

\begin{abstract}
In this paper, we introduce two alternative extensions of the classical univariate \emph{Value-at-Risk} (VaR) in a multivariate setting. The two proposed multivariate VaR are vector-valued measures with the same dimension as the underlying risk portfolio. 
The \emph{lower-orthant VaR} is constructed from level sets of multivariate distribution functions whereas the \emph{upper-orthant VaR} is constructed from level sets of multivariate survival functions.
Several properties have been derived. In particular, we show that these risk measures both satisfy the positive homogeneity and the translation invariance property. Comparison between univariate risk measures and components of multivariate VaR   are provided.  We also analyze how these measures are impacted by a change in marginal distributions, by a change in dependence structure and by a change in risk level.
%Interestingly, these results turn to be consistent with existing properties on univariate risk measures.
Illustrations are given in the class of  Archimedean copulas.
\end{abstract}

\end{frontmatter}

\section*{Introduction}

\noindent
During the last decades, researchers joined efforts to properly compare, quantify and manage
risk.  Regulators edict rules for bankers and insurers to improve
their risk management and to avoid crises, not always successfully as illustrated by recent
events. \\

\noindent
Traditionally, risk measures are thought of as mappings from a set of real-valued random
variables to the real numbers.  %\vspace{0.1cm}
However, it is often insufficient to consider a single real measure to quantify risks created by business activities, especially if the latter are affected by other external risk factors. Let us consider for instance the problem of solvency capital allocation for financial institutions with multi-branch businesses confronted to risks with specific characteristics. Under  Basel II and Solvency II, a bottom-up approach is used to estimate a  ``top-level''  solvency capital. This is done by using risk aggregation techniques who may capture risk mitigation or risk diversification effects. Then this global capital amount is re-allocated to each subsidiaries or activities for internal  risk management purpose (``top-down approach''). Note that the solvability of each individual branch  may strongly be affected by the degree of dependence amongst all branches. As a result, the capital allocated to each branch has to be computed in a multivariate setting where both  marginal effects and dependence between risks should be captured. In this respect, the ``Euler approach''  (e.g., see Tasche, \nocite{Tasche}2008) involving vector-valued risk measures has already been tested by risk management teams of some financial institutions.  \\

%\noindent
Whereas the previous risk allocation problem only involves internal risks associated with businesses in different  subsidiaries, the solvability of financial institutions could also be affected by external risks whose sources cannot be controlled. These risks may also be strongly heterogeneous in nature and difficult to diversify away. One can think for instance of systemic risk or contagion effects in a strongly interconnected system of financial companies.  As we experienced during the 2007-2009 crisis, the risks undertaken by some particular institutions may have significant impact on the solvability of the others. In this regard, micro-prudential regulation has been criticized because of its failure to limit the systemic risk within the system. This question has been dealt with recently by among others, Gauthier \emph{et al.} (2010)\nocite{Gauthier} and  Zhou (2010)\nocite{Zhou} who highlights the benefit of a  ``macro-prudential'' approach as an alternative solution to the existing ``micro-prudential'' one (Basel II) which does not take into account interactions between financial institutions.\\

\noindent
%From the years   $2000$ onward,
In the last decade, much research has been devoted to risk measures and many extensions to multidimensional settings have been investigated.
On theoretical grounds, Jouini \emph{et al.} (2004)\nocite{Jouini} proposes a class of set-valued coherent risk measures. Ekeland \textit{et al.}, (2012)\nocite{Galichon} derive a multivariate generalization of  Kusuoka's representation for coherent risk measures.
Unsurprisingly, the main difficulty regarding  multivariate generalizations  of risk measures is the fact that vector preorders are, in general, partial preorders. Then, what can be considered in a context of multidimensional portfolios as the analogous of a ``worst case''  scenario and a related ``tail distribution''? This is why several definitions of quantile-based risk measures are possible in a higher dimension.
%This is the first question we shall address by suggesting a suitable definition of quantiles for multi-risk portfolios.\\
%Several attempts have been made to provide a multidimensional generalization of the univariate quantile function.
For example,  Mass\'{e} and Theodorescu \nocite{Masse}(1994) defined multivariate quantiles as half-planes and  Koltchinskii (1997)\nocite{Koltchinskii} provided a general treatment of multivariate quantiles  as inversions of mappings.   Another approach is to use geometric quantiles  (see, for example, Chaouch \emph{et al.}\nocite{Chaouch}, 2009). Along with the geometric quantile, the notion of depth function  has been developed in recent years to characterize the quantile of multidimensional distribution functions (for further details see, for instance, Chauvigny \emph{et al.},  2011\nocite{Vero}).
 We refer to  Serfling (2002)\nocite{Serfling} for a large review on multivariate quantiles. When it turns to generalize  the \emph{Value-at-Risk} measure,  Embrechts and Puccetti (2006)\nocite{Embrechts}, Nappo and Spizzichino (2009\nocite{Nappo}), Pr\'{e}kopa (2012\nocite{Prekopa}) use the notion of quantile curve which is defined as the boundary of the upper-level set of a distribution function or the lower-level set of a survival function.
 % (formally introduced in Section \ref{Tail Value at Risk}).
 \\

\noindent
In this paper, we introduce two alternative extensions of the classical univariate \emph{Value-at-Risk} (VaR) in a multivariate setting. The proposed measures are based on the Embrechts and Puccetti (2006)'s\nocite{Embrechts} definitions of multivariate quantiles. We define the \emph{lower-orthant Value-at-Risk} at risk level $\alpha$ as the conditional expectation of the underlying vector of risks $\mathbf{X}$ given that the latter stands in the $\alpha$-level set of its distribution function. Alternatively, we define the \emph{upper-orthant Value-at-Risk} of $\mathbf{X}$ at level $\alpha$ as the  conditional expectation of $\mathbf{X}$ given that $\mathbf{X}$ stands in the $(1-\alpha$)-level set of its survival function. Contrarily to Embrechts and Puccetti (2006)'s approach, the extensions of \emph{Value-at-Risk} proposed in this paper are  real-valued vectors  with the same dimension as the considered portfolio of risks. This feature can be relevant from an operational point of view.\\

%These extensions may be useful to understand how solvency capital requirement (SCR) could be computed in a macro-prudential regulatory framework in which institutions can also be affected by risks undertaken by its competitors.
Several properties have been derived. In particular, we show that the \emph{lower-orthant Value-at-Risk} and the \emph{upper-orthant Value-at-Risk} both satisfy the positive homogeneity and the translation invariance property. We compare the components of these vector-valued measures with the univariate VaR of marginals.
We prove that the \emph{lower-orthant Value-at-Risk} (resp. \emph{upper-orthant Value-at-Risk}) turns to be more conservative (resp. less conservative) than the vector composed of  univariate VaR. We also analyze how these measures are impacted by a change in marginal distributions, by a change in dependence structure and by a change in risk level.
%Interestingly, these results turn to be consistent with existing properties on univariate risk measures.
In particular, we show that, for Archimedean families of copulas, the \emph{lower-orthant Value-at-Risk} and the \emph{upper-orthant Value-at-Risk} are both increasing   with respect to the risk level whereas their behavior is different with respect to the degree of dependence. In particular,  an increase of the dependence amongts risks tends to lower the \emph{lower-orthant Value-at-Risk} whereas it tends to widen the \emph{upper-orthant Value-at-Risk}.
%As a result,  contrary to the \emph{lower-orthant Value-at-Risk}, using the \emph{upper-orthant Value-at-Risk} as a measure of economic capital allows one to benefit from diversification effect since the sum of its components would be smaller than the sum of univ
%. This  suggests that this measure may be suitable for a capital allocation problems. Alternatively, the \emph{lower-orthant Value-at-Risk} could be apply to circumstances where risks cannot be diversify away.
In addition, these two measures may be useful for some applications where risks are heterogeneous in nature. Indeed, contrary to many existing approaches, no  arbitrary real-valued aggregate transformation is involved  (sum, min,  max,$\ldots$).\\

\noindent
The paper is organized as follows.  In Section \ref{Notation}, we  introduce some notations, tools and technical assumptions.
In Section  \ref{Multivariate generalization VAR}, we propose two multivariate extensions of  the \emph{Value-at-Risk} measure.
We study the properties of our multivariate VaR in terms of Artzner \emph{et al.} (1999)\nocite{Artzner}'s  invariance properties of risk measures (see Section  \ref{Invariance properties VAR}).  Illustrations in some  Archimedean copula cases  are presented in Section  \ref{Archimedean copula section}.  We also compare the components of these multivariate risk measures with the associated univariate  Value-at-Risk (see Section \ref{proprieta VAR bidimensionale}). The behavior of  our ${\rm VaR}$   with respect to a change in marginal distributions, a change in dependence structure and a change in risk level $\alpha$  is discussed respectively  in Sections \ref{stochastic order copulas var}, \ref{dependence structure copulas} and \ref{proprieta VAR PRD}.
   In the conclusion,  we discuss  open problems and  possible directions for  future work.

\section{Basic notions and  preliminaries}\label{Notation}

\noindent
In this section,  we first introduce some notation and tools  which will be used later on.

\subsection*{Stochastic orders}

From now on, let $Q_X(\alpha)$ be the univariate quantile function of a risk $X$ at level $\alpha \in (0,1)$.  More precisely, given an univariate continuous and strictly monotonic loss distribution function $F_X$,  $Q_X(\alpha)= F_X^{-1}(\alpha)$, $\forall\, \alpha \, \in \, (0,1)$.   We recall here the definition and some properties of  useful univariate and multivariate stochastic  orders.

\begin{Definition}[Stochastic dominance order] \label{def st order}
Let $X$ and $Y$ be two random variables. Then $X$ is said to be smaller than $Y$ in stochastic dominance, denoted as $\, X \preceq_{st} Y$,  if the inequality  $Q_X(\alpha) \leq Q_Y(\alpha)$ is satisfied for all $\alpha \in (0, 1).$
\end{Definition}

\begin{Definition}[Stop-loss order] \label{def sl  order}
\begin{sloppypar}
Let $X$ and $Y$ be two random variables. Then $X$ is said to be smaller than $Y$ in the  stop-loss order, denoted as $\, X \preceq_{sl} Y$,  if  for all $t \in \Real,$  ${ \E[(X- t)_+] \leq \E[(Y- t)_+]},$ with $x_+ :=\max\{x,0\}$.\end{sloppypar}
\end{Definition}

\begin{Definition}[Increasing convex order] \label{def icx order}
Let $X$ and $Y$ be two random variables. Then $X$ is said to be smaller than $Y$ in the  increasing  convex order, denoted as $\, X \preceq_{icx} Y$,  if  $\,\E[f(X)] \leq \E[f(Y)],$ for all non-decreasing  convex function $f$ such that the expectations exist.
\end{Definition}

\noindent
The stop-loss order and the increasing convex order are equivalent (see Theorem 1.5.7 in  M{\"u}ller and   Stoyan, 2001\nocite{Mullerbook}).   Note  that stochastic dominance order implies stop-loss order.  For more details about stop-loss order we refer the interested reader to  M{\"u}ller  (1997\nocite{Muller}).\\

%Moreover, a sufficient condition for the stop-loss order   is the \textit{dangerousness order relation} as stated in the following lemma.
%\begin{Lemma}[Ohlin, 1969\nocite{Ohlin}]\label{dangerousness order}
%\begin{sloppypar}
% Let $X$ and $Y$ be random variables with
%finite means such that ${\E[X] \leq \E[Y]}$, and there exists  some real number $c$ such that $F_X(x)\leq F_Y(x),  \mbox{ for all } x < c$   and
%$F_X(x) \geq F_Y(x),  \mbox{ for all } x  \geq c$. Then $X$ precedes $Y$  in dangerousness order, written $X \preceq_D Y$,
%and this implies the stop-loss order  $X \preceq_{sl} Y$.
%\end{sloppypar}
%\end{Lemma}
%
%\noindent
%According to B\"{u}hlmann's terminology, when $X \preceq_D Y$  the distribution function  $F_Y$ is said to be more dangerous than $F_X$.  This terminology  is essentially related to the variability of the random variables $X$ and $Y$ (see Section 3.4.2.2 in  Denuit et al., 2005\nocite{Denuit}).  For further details, the reader is referred to    B\"{u}hlmann \emph{et  al.} (1977\nocite{Buhlmann}).

Finally,  we introduce the  definition of supermodular function and supermodular order  for  multivariate random vectors.

\begin{Definition}[Supermodular function] \label{def sm  fucntion}
A function  $f: \Real^d \rightarrow \Real$ is said to be  supermodular if
for any $\textbf{x}, \textbf{y} \in \Real^d $  it satisfies
$$f(\textbf{x})+ f(\textbf{y}) \leq f(\textbf{x} \wedge \textbf{y}) + f(\textbf{x} \vee \textbf{y}),$$
where the operators $\wedge$ and $\vee$ denote coordinatewise minimum and maximum respectively.
\end{Definition}

\begin{Definition}[Supermodular order] \label{def sm  order}
Let $\textbf{X}$ and $\textbf{Y}$ be two  $d-$dimensional random vectors such that $\, \E[f(\textbf{X})] \leq \E[f(\textbf{Y})],\,$  for all supermodular functions $f: \Real^d \rightarrow \Real$, provided the expectation exist.  Then  $\textbf{X}$ is said to be smaller than $\textbf{Y}$ with respect to the supermodular order (denoted  by $\, \textbf{X} \preceq_{sm} \textbf{Y}$).
\end{Definition}\vspace{0.1cm}

\noindent
This will be  a key tool to analyze the impact of dependence on our multivariate risk measures.

\subsection*{Kendall distribution function}
\vspace{0.1cm}

Let $\textbf{X}= (X_1, \ldots, X_d)$ be a $d-$dimensional random vector, $d\geq 2$.  As we will see later on, our study of multivariate risk measures strongly relies on the key concept of  Kendall distribution function (or multivariate probability integral transformation),  that is, the distribution function of the  random variable $F(\textbf{X})$,  where $F$ is the multivariate distribution of random vector $\textbf{X}$.   From now on, the Kendall distribution will be denoted by $K$, so that $K(\alpha) = \Pbb[F(\textbf{X})\leq \alpha]$, for $\alpha \in [0,1]$.  We also denote by $\overline{K}(\alpha)$ the survival distribution function of  $F(\textbf{X})$, i.e., $\overline{K}(\alpha)= \Pbb[F(\textbf{X}) > \alpha]$.
For more details on the  multivariate probability integral transformation, the interested reader is referred  to Cap{\'e}ra{\`a} \emph{et al.},  (1997\nocite{kendalorder}),  Genest and Rivest (2001)\nocite{Rivest}, Nelsen \textit{et al.} (2003)\nocite{Nelsen}, Genest and Boies (2003)\nocite{Genest1},  Genest \textit{et al.} (2006)\nocite{Genest2} and  Belzunce \emph{et al.}  (2007).
\\

\noindent
In contrast to the univariate case, it is not generally true that the
distribution function $K$ of $F(\textbf{X})$ is uniform on $[0, 1]$, even when $F$ is continuous.   Note also that it is not  possible to characterize the joint distribution  $F$ or reconstruct it from the knowledge of $K$ alone, since the latter does not contain any information about
the marginal distributions $F_{X_{1}}, \ldots, F_{X_{d}}$  (see Genest and Rivest, 2001\nocite{Rivest}). Indeed, as a consequence of Sklar's Theorem,  the Kendall distribution only depends on the dependence structure or the copula function $C$ associated with $\textbf{X}$ (see Sklar, 1959\nocite{Sklar}). Thus, we also have $K(\alpha) = \Pbb[C(\textbf{U})\leq \alpha]$ where $\textbf{U}= (U_1, \ldots, U_d)$ and  $U_1= F_{X_{1}}(X_1), \ldots, U_d= F_{X_{d}}(X_d)$.   \\

\noindent
  Furthermore:\vspace{0.2cm}
\begin{itemize}
\item For a $d-$dimensional  random vector $\textbf{X}= (X_1, \ldots, X_d)$ with copula $C$, the Kendall distribution function  $K(\alpha)$ is linked  to the  Kendall's tau correlation coefficient via: $\tau_C =   \frac{2^d\, \E[C(\textbf{U})]-1}{2^{d-1}-1}$, for $d \geq 2$  (see  Section 5 in Genest and Rivest, 2001\nocite{Rivest}).\vspace{0.1cm}
\item \begin{sloppypar} The Kendall distribution can be obtain explicitly in the case of multivariate Archimedean copulas with generator\footnote{Note that $\phi$ generates a $d-$dimensional Archimedean copula if and only if its inverse $\phi^{-1}$ is a $d-$ monotone on $[0, \infty)$ (see Theorem 2.2 in McNeil and Ne\v{s}lehov\'{a}, 2009). \nocite{McNeil_Neslehova}}  $\phi$, i.e., $C(u_1, \ldots, u_d) = \phi^{-1}\left(\phi(u_1)+\cdots+\phi(u_d)\right)$ for all $(u_1, \ldots, u_d)\in [0,1]^d.$ Table \ref{kendall classiche} provides the expression of Kendall distributions associated with Archimedean, independent and comonotonic  $d-$dimensional random vectors (see Barbe \emph{et al.}, 1996\nocite{Barbe}). Note that the Kendall distribution is uniform for comonotonic random vectors.\end{sloppypar}
\begin{table}[!ht]
 \begin{minipage}[h!]{0.99\linewidth}\centering
\begin{tabular}{| c | c| c | c | c |c |}
 \hline \centering
      {\Large\strut} Copula   &  Kendall distribution  $K(\alpha)$ \\  \hline
       {\huge\strut} Archimedean  case &    $\alpha + \, \sum_{i=1}^{d-1} \, \frac{1}{i!}\left(-\phi(\alpha)\right)^{i} \, \left(\phi^{-1}\right)^{(i)}\left(\phi(\alpha)\right)$   \\
       		%f_{i-1}(\alpha)$      \\ \vspace{0.02cm}
             %Counter-monotonic case &  $1$ \\  \vspace{0.02cm}
       {\huge\strut} Independent  case   &    $\alpha +\alpha\, \sum_{i=1}^{d-1} \left(\frac{\ln(1/\alpha)^i}{i!}\right)$\\
      Comonotonic case & $\alpha$  \\
      \hline
   \end{tabular}
   \caption{{\small Kendall distribution in some classical  $d-$dimensional dependence structure.}}
   \label{kendall classiche}
      \end{minipage}
      \end{table}\vspace{0.2cm}
%       In Table   \ref{kendall classiche},   $f_{i}(\alpha)$ stands for $\frac{\rmd^{i-1} \phi^{-1}(t)}{\rmd t^{i+1}}$, evaluated at  $t= \phi(\alpha)$.
       For further details the interested reader is referred to Section 2 in Barbe \emph{et al.} (1996\nocite{Barbe}) and  Section 5 in Genest and Rivest (2001\nocite{Rivest}).    For instance, in the bivariate case, the Kendall distribution function is equal to  $\alpha-\frac{\phi(\alpha) }{\phi'(\alpha)},$ $\alpha\in(0,1),$ for Archimedean copulas with differentiable generator $\phi.$ It is equal to $ \alpha\left(1-\ln(\alpha)\right),$ $\alpha\in(0,1)$ for the bivariate independence copula and to $1$ for the counter-monotonic bivariate copula.
 \begin{sloppypar}
 \item  It holds that $\alpha \leq K(\alpha) \leq 1$, for all $\alpha \, \in (0,1),$ i.e., the graph of the  Kendall distribution function is above the first diagonal (see Section 5 in  Genest and Rivest, 2001\nocite{Rivest}).  This is equivalent to state that, for any random vector $\textbf{U}$ with copula function $C$ and uniform marginals,
  $C(\textbf{U})  \preceq_{st}C^{\text{c}}(\textbf{U}^{\text{c}}) $ where  $\textbf{U}^{\text{c}}=(U_{1}^{\text{c}},\ldots, U_{d}^{\text{c}})$ is a comonotonic random vector with  copula function $C^{\text{c}}$ and uniform marginals.
% $C^{\text{cc}}(\textbf{U}^{\text{cc}}) \preceq_{st} C(\textbf{U})  \preceq_{st}C^{\text{c}}(\textbf{U}^{\text{c}}) $ where  $\textbf{U}^{\text{cc}}=(U_{1}^{\text{cc}},\ldots, U_{d}^{\text{cc}})$ (resp. $\textbf{U}^{\text{c}}=(U_{1}^{\text{c}},\ldots, U_{d}^{\text{c}})$) is a counter-monotonic (resp. comonotonic) random vector with  copula function $C^{\text{cc}}$ (resp. $C^{\text{c}}$) and uniform marginals.
 \end{sloppypar}
\end{itemize}
 \begin{sloppypar}
\noindent
This last property suggests that  when the level of dependence between $X_1, \ldots, X_d$  increases, the Kendall distribution also  increases in some sense. The following result, using definitions of stochastic orders described above,  investigates  rigorously this intuition. \end{sloppypar}

\begin{Proposition}\label{super modal order}
Let $\textbf{U}= (U_1, \ldots, U_d)$ {\rm(}resp. $\textbf{U}^*= (U_{1}^*, \ldots, U_{d}^*)${\rm)} be a  random vector  with  copula $C$ (resp. $C^*$) and uniform marginals.
\begin{center}
If $ \, \, \textbf{U} \preceq_{sm}  \textbf{U}^*, \, \,$
 then  $ \, \,C(\textbf{U}) \preceq_{sl} C^*(\textbf{U}^*).$
\end{center}
\end{Proposition}

\begin{sloppypar}
\noindent
\emph{Proof:} Trivially, $\textbf{U} \preceq_{sm} \textbf{U}^*   \Rightarrow  C(\textbf{u}) \leq C^*(\textbf{u})$, for all $\textbf{u} \in  [0,1]^d$  (see Section 6.3.3  in Denuit \emph{et al.}, 2005\nocite{Denuit}).   Let $f: [0,1] \rightarrow \Real$  be a  non-decreasing   and convex function. It holds that $f(C(\textbf{u})) \leq f(C^*(\textbf{u}))$, for all $\textbf{u} \in  [0,1]^d$, and  $\E[f(C(\textbf{U}))]\leq \E[f(C^*(\textbf{U}))]$. Remark that since $C^*$ is   non-decreasing  and supermodular and $f$ is  non-decreasing and convex  then $f \circ  C^*$ is a   non-decreasing and supermodular function (see Theorem 3.9.3  in  M{\"u}ller and   Stoyan, 2001\nocite{Mullerbook}). Then, by assumptions, $\E[f(C(\textbf{U}))] \leq \E[f(C^*(\textbf{U}))] \leq  \E[f(C^*(\textbf{U}^*))]$. This implies   $C(\textbf{U}) \preceq_{sl} C^*(\textbf{U}^*).$  Hence the result. $\Box$\\ \end{sloppypar}
%\vspace{0.15cm}
\begin{sloppypar}
\noindent
From Proposition \ref{super modal order},  we remark that $\textbf{U}\preceq_{sm} \textbf{U}^*$  implies an  ordering relation between  corresponding Kendall's tau :  $\tau_{C} \leq \tau_{C^*}$.  Note that the supermodular order between $\textbf{U}$ and $\textbf{U}^*$
%(i.e.,  $C(\textbf{u}) \leq C^*(\textbf{u})$, for all $\textbf{u} \in  [0,1]^d$)
does not necessarily yield the stochastic dominance order between $C(\textbf{U})$ and $C^*(\textbf{U}^*)$ (i.e.,  ${C(\textbf{U}) \preceq_{st} C^*(\textbf{U}^*)}$ does not hold in general).  For a bivariate counter-example, the interested reader is referred to,  for instance, Cap{\'e}ra{\`a} \emph{et al.}   (1997\nocite{kendalorder}) or Example 3.1 in Nelsen \emph{et al.} (2003).\\
\end{sloppypar}

\noindent
Let us now focus on some classical families of bivariate Archimedean copulas. In Table  \ref{kendall achimedeans}, we obtain analytical expressions of the Kendall distribution function for Gumbel, Frank, Clayton and Ali-Mikhail-Haq families.
\begin{table}[!ht]
 \begin{minipage}[h!]{0.99\linewidth}\centering
\begin{tabular}{| c | c| c | c | c |c |}
 \hline \centering
     {\large\strut}  Copula   &  $\theta \in $ & Kendall distribution  $K(\alpha, \theta)$\\  \hline \vspace{0.02cm}
      {\large\strut}  Gumbel  & $[1, \infty )$ & $\alpha\left(1-{\frac {1 }{\theta}\ln\alpha}\right)$ \\ \vspace{0.02cm}
      {\large \strut}  Frank    & $(-\infty,\infty)  \setminus  \{0\}$ & ${ \alpha+ \frac{1}{\theta}\left(1-{\rm e}^{\theta\alpha}\right)\ln \left( {\frac {{1-{\rm e}^{-\theta\,\alpha}}}{{1-{\rm e}^{-\theta}}}} \right)}$ \\ \vspace{0.02cm}
      {\large \strut}  Clayton  & $[-1,\infty)  \setminus  \{0\}$ & $ \alpha\left(1+ \frac{1}{\theta}{\left(1-\alpha^{\theta}\right)}\right)$ \\ \vspace{0.02cm}
    % {\Large\strut}   Gumbel-Barnett &  $\theta  \in (0,1]$ & $\alpha+{\frac {\ln  \left( 1-\ln  \left( \alpha \right) \theta
 % \right) \alpha\, \left( 1-\ln  \left( \alpha \right) \theta \right) }{\theta}}$  \\  \vspace{0.02cm}
 {\large \strut}  Ali-Mikhail-Haq &
%  $[-1,1)$ & $1+\frac{\alpha}{\theta-1}+\left(\frac{\theta\alpha}{\theta-1}-1\right)\left(\ln\left(\theta(\alpha+1)-1\right) - \ln\alpha\right) $ \\  \hline
 $[-1,1)$ & $\frac {\alpha\,  -1+\theta+   (1- \theta + \theta \alpha) (\ln  \left( 1-\theta+\theta\,\alpha \right) + \ln\alpha) }{\theta-1}$ \\  \hline
   \end{tabular}\vspace{0.1cm}
    \caption{{\small Kendall distribution in some bivariate  Archimedean cases. \vspace{0.2cm}}}
   \label{kendall achimedeans}
      \end{minipage}
\end{table}
\begin{Remark}\label{theta e dimensione}
Bivariate Archimedean copula can be extended to $d$-dimensional copulas with $d> 2$ as far as the generator $\phi$ is a $d$-monotone function on $[0, \infty)$  (see McNeil and Ne\v{s}lehov\'{a}, 2009\nocite{McNeil_Neslehova} for more details). For the $d$-dimensional Clayton copulas, the underlying dependence parameter must be such that $ \theta > - \frac{1}{d-1}$  (see Example 4.27 in Nelsen, 1999\nocite{NelsenLibro}). Frank copulas can be extended to $d$-dimensional copulas for $\theta>0$ (see Example 4.24 in Nelsen, 1999\nocite{NelsenLibro}).
\end{Remark}

\begin{figure}[h!]
\begin{minipage}[b]{0.5\linewidth}\centering
\includegraphics[width=8.2cm]{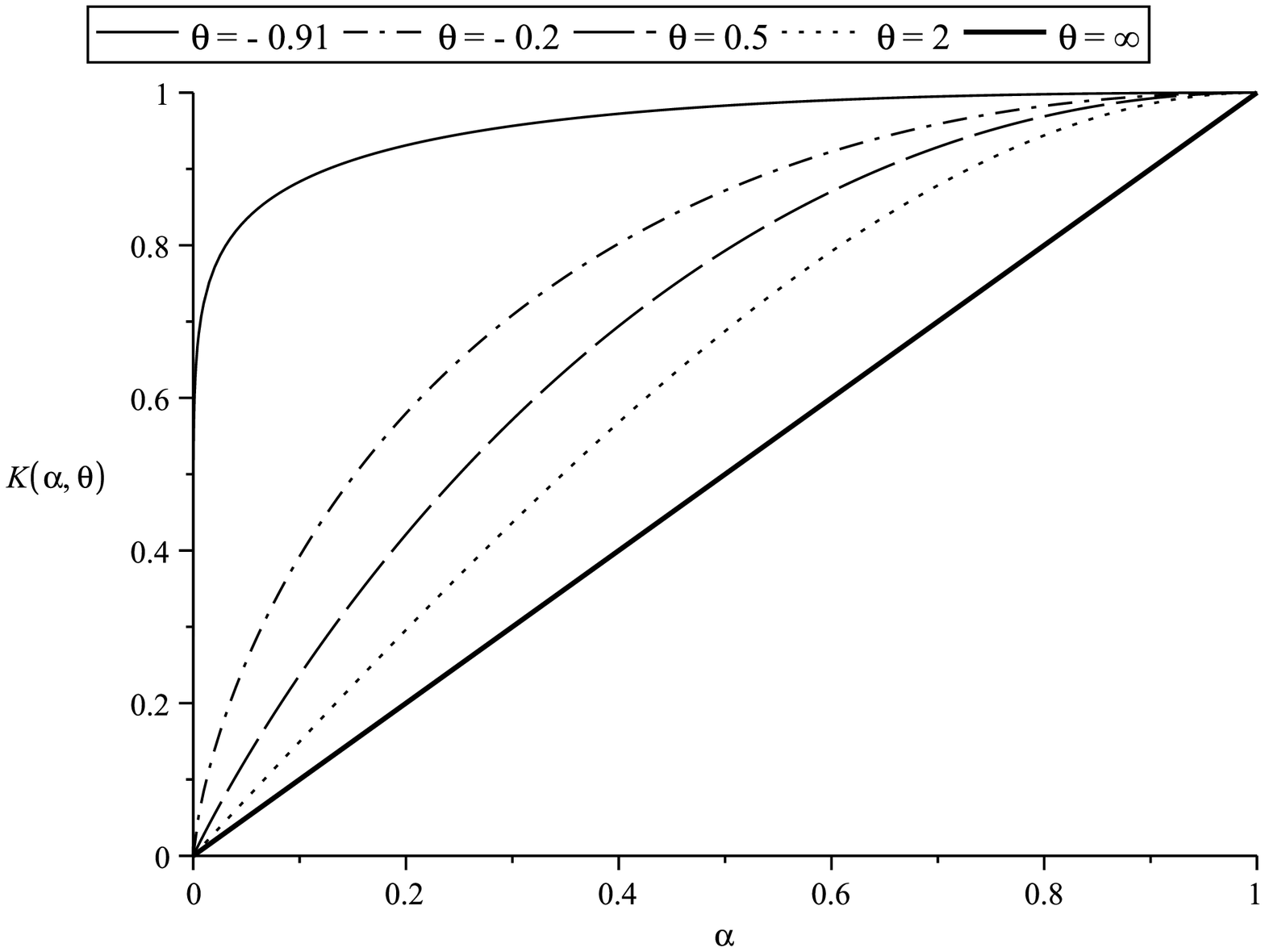}
\end{minipage}
\begin{minipage}[b]{0.4\linewidth}\centering
\includegraphics[width=8.4cm]{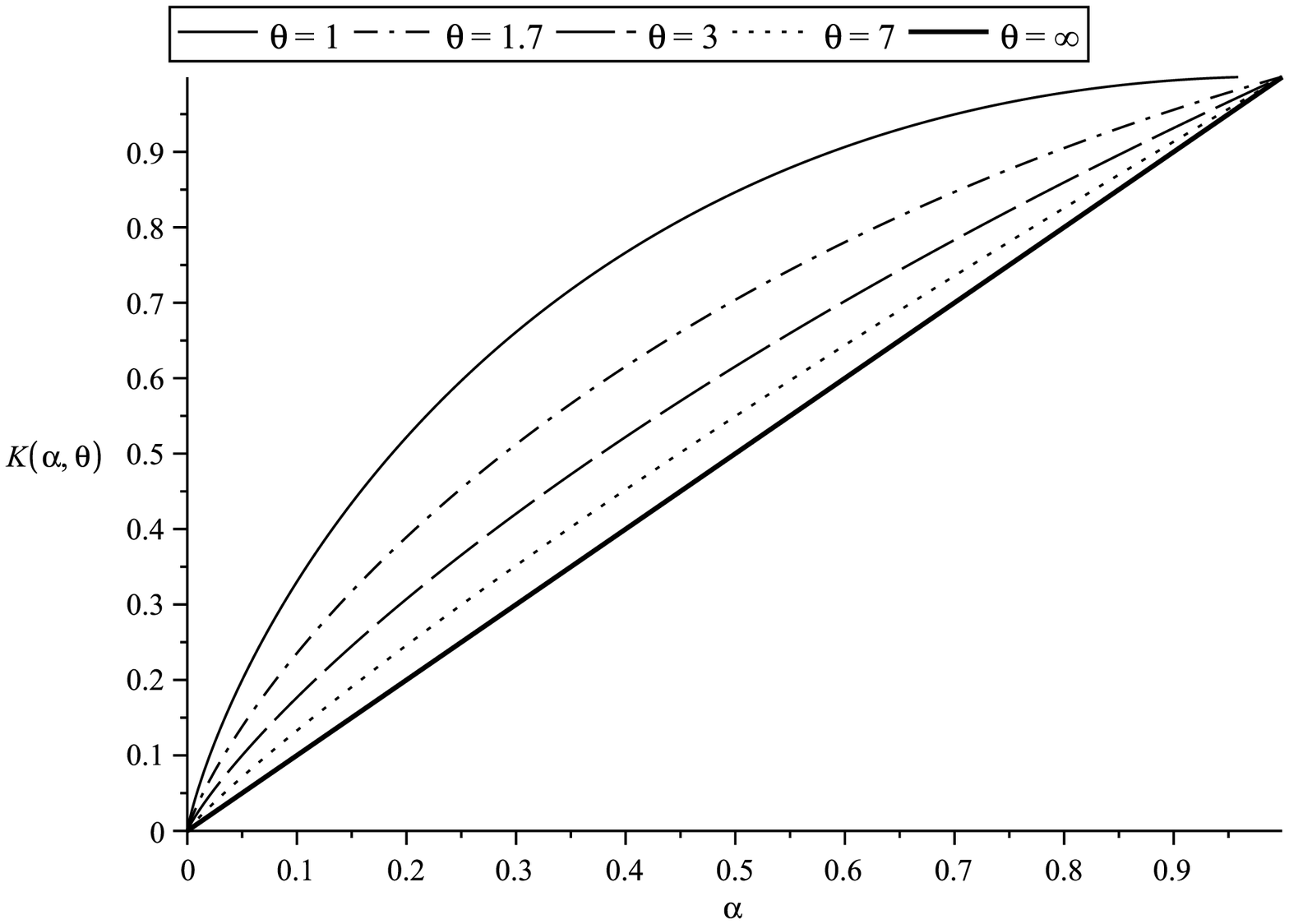}
\end{minipage}
 \caption{{\small Kendall distribution $K(\cdot, \theta)$ for different values of $ \theta$ in the Clayton copula case (left) and the Gumbel copula case (right). The dark full line represents the  first diagonal and it corresponds to the comonotonic case.}}\label{Kendall alpha theta gumbel}
\end{figure}

\noindent
Note that parameter $\theta$ governs the level of dependence amongst components of the underlying random vector. Indeed, it can be shown that, for all Archimedean copulas in Table \ref{kendall achimedeans}, an increase of $\theta$ yields an increase of dependence in the sense of the supermodular order, i.e., $\theta \leq \theta^* \Rightarrow \textbf{U} \preceq_{sm}\textbf{U}^*$  (see further examples in Joe, 1997 and   Wei and Hu, 2002\nocite{Wei}). Then, as a consequence of Proposition \ref{super modal order}, the following comparison result holds
 \begin{equation*}
  \theta \leq \theta^* \Rightarrow C(\textbf{U}) \preceq_{sl} C^*(\textbf{U}^*).
  %\Rightarrow K(\alpha, \theta) \geq K^*(\alpha, \theta^*)$,
 \end{equation*}
In fact, a stronger comparison result can be derived for Archimedean copulas of Table  \ref{kendall achimedeans}, as shown in the following remark.
\begin{Remark}\label{St_Kendall_Archimedean}
 \begin{sloppypar}
 \noindent
For  copulas in Table  \ref{kendall achimedeans},  one can check that $\frac{\partial K(\alpha, \theta)}{\partial \theta} \leq 0$, for all $\alpha \in (0,1)$.
This means that, for these classical examples, the associated Kendall distributions actually increase  with respect to the  stochastic dominance order when the dependence parameter $\theta$ increases, i.e.,
 \begin{equation*}
 \label{eq_St_Dominance}
{\theta \leq \theta^*  \Rightarrow C(\textbf{U}) \preceq_{st} C^*(\textbf{U}^*)}.
 \end{equation*}
In order to illustrate this property  we plot  in Figure \ref{Kendall alpha theta gumbel} the Kendall distribution function $K(\cdot, \theta)$   for different choices of  parameter $\theta$ in the bivariate Clayton copula case and in the bivariate Gumbel copula case.  \end{sloppypar}
\end{Remark}

\section{Multivariate generalization of the Value-at-Risk measure}\label{Multivariate generalization VAR}

From the usual definition in the univariate setting, the \emph{Value-at-Risk} is the minimal amount of the loss which accumulates a probability $\alpha$ to the left tail and $1 - \alpha$ to the right tail. Then, if $F_X$ denotes the cumulative distribution function associated with risk $X$ and $\overline{F}_X$ its associated survival function, then
$$
\VAR_\alpha(X) := \inf \left\{x\in \Real : F_X(x) \geq \alpha \right\}% := Q_X(\alpha), \quad \forall\, \, \, \alpha \, \in \, (0,1).
$$
and equivalently,
$$
\VAR_\alpha(X) := \inf \left\{x\in \Real : \overline{F}_X(x) \leq 1-\alpha \right\}.% := Q_X(\alpha), \quad \forall\, \, \, \alpha \, \in \, (0,1).
$$

%we know that the quantile function $Q_X(\alpha)$ provides a point which accumulates a probability $\alpha$ to the left tail and $1 - \alpha$ to the right tail.   The univariate quantile function  $Q_X$ is used in risk theory to define an univariate risk measure: the \emph{Value-at-Risk}. This measure is defined as
%$$
%\rm{VaR}_\alpha(X)= \inf \left\{x\in \Real : F_X(x) \geq \alpha \right\} := Q_X(\alpha), \quad \forall\, \, \, \alpha \, \in \, (0,1).
%$$
Consequently, the classical univariate VaR can be viewed as  the boundary of the set $\left\{x\in \Real :\right.$ $\left. F_X(x) \geq \alpha \right\}$ or, similarly, the boundary of the set $\left\{x\in \Real : \overline{F}_X(x) \leq 1-\alpha \right\}$.\\

 This idea can be easily extended in higher dimension, keeping in mind that the two previous sets are different in general as soon as $d\geq 2$. We propose a multivariate generalization of \emph{Value-at-Risk} for a portfolio of $d$ dependent risks. As a starting point, we consider  Definition 17 in  Embrechts and Puccetti (2006\nocite{Embrechts}). They suggest to define the multivariate lower-orthant  Value-at-Risk at probability
level $\alpha$, for a increasing function $\underline{G}$ : $\Real^d \rightarrow [0,1]$, as the boundary of its $\alpha$--upper-level set, i.e., $\partial \{\textbf{x} \in \mathbb{R}^{d} : \underline{G}(\textbf{x}) \geq \alpha\}$ and  analogously, the multivariate upper-orthant   Value-at-Risk, for a decreasing function
$\overline{G}$ : $\Real^d \rightarrow [0,1]$, as the  boundary of its $(1-\alpha)$--lower-level set, i.e., $\partial \{\textbf{x} \in \mathbb{R}^{d} : \overline{G}(\textbf{x})  \leq 1- \alpha\}$. \\

Note that the generalizations of \emph{Value-at-Risk} by Embrechts and Puccetti (2006\nocite{Embrechts}) (see also Nappo and Spizzichino, 2009\nocite{Nappo}; Tibiletti, 1993\nocite{Tibiletti3}) are represented by an infinite number of points  (an hyperspace of dimension $d-1$,  under some regularly conditions  on the functions $\underline{G}$ and $\overline{G}$).  This  choice  can be unsuitable  when we face real risk management problems.  Then, % in Sections \ref{Tail Value at Risk} and \ref{Tail Value at Risk upper}
we propose more parsimonious and synthetic versions of the Embrechts and Puccetti (2006)'s  measures.  In particular in our propositions,  instead of considering the whole hyperspace $\partial \{\textbf{x}   : \underline{G}(\textbf{x}) \geq \alpha\}$  (or $\partial \{\textbf{x} : \overline{G}(\textbf{x})  \leq 1- \alpha\}$)  we only focus on the particular point in $\mathbb{R}_+^d$  that matches  the conditional expectation of $\textbf{X}$ given that $\textbf{X}$ stands in this set.   This means that our measures are  real-valued vectors  with the same dimension as the considered portfolio of risks.\\

%The latter feature could be relevant on practical grounds.
In addition, to be consistent with the univariate definition of  $\rm{VaR}$, we choose $\underline{G}$ (resp. $\overline{G}$) as the  $d-$dimensional loss distribution function $F$ (resp. the survival distribution function $\overline{F}$) of the risk portfolio. This allows to capture information coming both from the marginal distributions and from the multivariate dependence structure, without  using  an arbitrary real-valued aggregate transformation (for more details see Introduction). \\

In analogy with the Embrechts and Puccetti's notation we will denote    $\underline{{\rm VaR}}$ our multivariate lower-orthant  Value-at-Risk
%(see Section \ref{Tail Value at Risk})
 and $\overline{{\rm VaR}}$ the  upper-orthant one.\\% (see Section \ref{Tail Value at Risk upper}).\\

\begin{sloppypar}
\noindent
In the following,  we will consider  non-negative  absolutely-continuous random vector\footnote{We restrict ourselves to $\mathbb{R}^d_+$ because, in our applications, components of $d-$dimensional vectors correspond to random losses and are  then  valued in  $\mathbb{R}_+$.}   ${\textbf{X}= (X_1,  \ldots, X_d)}$   (with respect to   Lebesgue measure $\lambda$ on $\mathbb{R}^d$) with  partially  increasing multivariate distribution function\footnote{A function $F(x_1,\ldots, x_d)$ is partially   increasing on $\mathbb{R}^{d}_{+} \setminus (0,\ldots, 0)$ if the functions of one variable $g(\cdot)= F(x_1, \ldots, x_{j-1}, \cdot, x_{j+1}, \ldots,  x_d)$  are  increasing. About properties of partially   increasing multivariate distribution functions we refer the interested reader to  Rossi (1973\nocite{Rossi}),  Tibiletti  (1991\nocite{Tibiletti4}).}   $F$   and such that ${\E(X_i)< \infty},$ for $i=1, \ldots, d$. These conditions will be called \emph{regularity conditions}. \end{sloppypar}
\vspace{0.17cm}

\noindent
However,  extensions of our results in the case of  multivariate distribution function on the entire space $\mathbb{R}^{d}$ or  in the presence of plateau in the graph of $F$ are possible.  Starting from these considerations, we introduce here a multivariate  generalization of the \rm{VaR}  measure.

\begin{Definition}[Multivariate lower-orthant Value-at-Risk]\label{VAR multivariate}
Consider a random vector $\textbf{X}=(X_1,\ldots, X_d)$ with distribution function $F$  satisfying  the  regularity conditions.  For $\alpha \in (0, 1)$, we define  the multidimensional lower-orthant  Value-at-Risk  at probability  level $\alpha$ by
\begin{equation*}
\underline{{\rm VaR}}_\alpha(\textbf{X})=\E[\textbf{X}|\, \textbf{X} \in \partial \underline{L}(\alpha) ]=\left(
                                              \begin{array}{ll}
                                                \E[\,X_1\,|\,\textbf{X} \in \partial \underline{L}(\alpha)\,] \vspace{0.04cm} \\
                                                \quad \quad \quad \quad   \vdots  \vspace{0.04cm} \\
                                                \E[\,X_d\,|\,\textbf{X} \in \partial \underline{L}(\alpha)\,]
                                              \end{array}
                                            \right).
\end{equation*}
where $\partial \underline{L}(\alpha)$ is the boundary of the set $\underline{L}(\alpha):= \{\textbf{x} \in \mathbb{R}^{d}_+ : F(\textbf{x})  \geq \alpha\}$.
Under the regularity conditions, $\partial \underline{L}(\alpha)$ is the $\alpha$-level set of $F$, i.e., $\partial \underline{L}(\alpha)=\{\textbf{x} \in \mathbb{R}^{d}_+ : F(\textbf{x})  = \alpha\}$ and the previous definition can be restated as
\begin{equation*}
\underline{{\rm VaR}}_\alpha(\textbf{X}) =\E[\textbf{X}|\,F(\textbf{X}) = \alpha]=\left(
                                              \begin{array}{ll}
                                                \E[\,X_1\,|\,F(\textbf{X})= \alpha\,]\vspace{0.04cm} \\
                                                  \quad \quad \quad  \quad  \vdots  \vspace{0.04cm} \\
                                                \E[\,X_d\,|\,F(\textbf{X})= \alpha\,]
                                              \end{array}
                                            \right).
\end{equation*}
\vspace{0.05cm}
%where $\partial \underline{L}(\alpha)$ is the $\alpha$-level set of $F$, i.e., $\partial \underline{L}(\alpha)= \partial\{\textbf{x} \in \mathbb{R}^{d}_+ : F(\textbf{x})  \geq \alpha\}$.
\end{Definition} \vspace{0.15cm}

\noindent
Note that, under the \emph{regularity conditions},   $\partial \underline{L}(\alpha)=  \{\textbf{x} \in \mathbb{R}^{d}_{+}: F(\textbf{x}) = \alpha\}$  has  Lebesgue-measure zero in $\mathbb{R}^{d}_{+}$ (e.g., see Property 3 in Tibiletti, 1990\nocite{Tibiletti1990}). Then we make sense of  Definition  \ref{VAR multivariate}  using the limit procedure in Feller (1966\nocite{Feller}), Section 3.2:
\begin{multline}\label{limit procedure}
\E[\,X_i \,|\, F(\textbf{X}) =\alpha\,]=  \lim_{h \rightarrow 0}\, \E[\,X_i  \,|\,\alpha <F(\textbf{X}) \leq  \alpha +h \,] \\ =
 \lim_{h \rightarrow 0}\,  \frac{   \int_{Q_{X_i}(\alpha)}^\infty  x \left( \int_{\alpha}^{\alpha +h }  f_{(X_i,F(\textbf{X}))}(x,y) \, \rmd y\right)  \rmd x}{ \int_{\alpha}^{\alpha +h }  f_{F(\textbf{X})}(y) \,  \rmd y},
\end{multline}
for $i= 1,\ldots, d$.

\noindent
Dividing numerator and denominator  in \eqref{limit procedure} by $h$, we obtain, as $h \rightarrow 0$
\begin{equation}\label{VAR multivariate formula}
\E[X_i \, | F(\textbf{X}) =\alpha]%=  \underline{{\rm VaR}}^i_\alpha(\textbf{X}) =
=  \frac{ \int_{Q_{X_i}(\alpha)}^\infty  x\,   f_{(X_i,F(\textbf{X}))}(x, \alpha) \, \rmd x} {K'(\alpha)},\end{equation}
for $i= 1,\ldots, d$,  where $K'(\alpha) = \frac{\rmd K(\alpha)}{\rmd \alpha}$ is the Kendall distribution density function. This procedure gives a rigorous sense to our  $\underline{{\rm VaR}}_\alpha(\textbf{X})$ in Definition \ref{VAR multivariate}.  Remark that the existence of $f_{(X_i,F(\textbf{X}))}$ and $K'$ in \eqref{VAR multivariate formula} is guaranteed by the  \emph{regularity conditions} (for further details, see  Proposition 1 in  Imlahi \emph{et al.}, 1999\nocite{Imlahi} or Proposition 4 in Chakak and Ezzerg, 2000\nocite{Chakak}).\\

In analogy with Definition \ref{VAR multivariate}, %and using again the idea in Definition 17 in  Embrechts and Puccetti (2006\nocite{Embrechts}),
we now  introduce another possible generalization of the \rm{VaR}  measure based on the survival distribution function.

\begin{Definition}[Multivariate upper-orthant Value-at-Risk]\label{VAR multivariate upper}
Consider a random vector $\textbf{X}=(X_1,\ldots, X_d)$ with survival distribution $\overline{F}$ satisfying  the  regularity conditions.  For $\alpha \in (0, 1)$, we define  the multidimensional upper-orthant  Value-at-Risk  at probability  level $\alpha$ by
\begin{equation*}
\overline{{\rm VaR}}_\alpha(\textbf{X})=\E[\textbf{X}|\, \textbf{X} \in \partial \overline{L}(\alpha) ]=\left(
                                              \begin{array}{ll}
                                                \E[\,X_1\,|\,\textbf{X} \in \partial \overline{L}(\alpha)\,] \vspace{0.04cm} \\
                                                \quad \quad \quad \quad   \vdots  \vspace{0.04cm} \\
                                                \E[\,X_d\,|\,\textbf{X} \in \partial \overline{L}(\alpha)\,]
                                              \end{array}
                                            \right).
\end{equation*}
where $\partial \overline{L}(\alpha)$ is the boundary of the set $\overline{L}(\alpha):=  \{\textbf{x} \in \mathbb{R}^{d}_+ : \overline{F}(\textbf{x})  \leq 1 - \alpha\}$.
Under the regularity conditions, $\partial \overline{L}(\alpha) $ is the $(1-\alpha)$-level set of $\overline{F}$, i.e., $\partial \overline{L}(\alpha) = \{\textbf{x} \in \mathbb{R}^{d}_+ : \overline{F}(\textbf{x})  = 1 - \alpha\}$ and the previous definition can be restated as
\begin{equation*}
\overline{{\rm VaR}}_\alpha(\textbf{X}) =\E[\textbf{X}|\, \overline{F}_{\textbf{X}}(\textbf{X})=1-\alpha ]=\left(
                                              \begin{array}{ll}
                                                \E[\,X_1\,|\,  \overline{F}(\textbf{X})=1-\alpha\,]\vspace{0.04cm} \\
                                                  \quad \quad \quad  \quad  \vdots  \vspace{0.04cm} \\
                                                \E[\,X_d\,|\, \overline{F}(\textbf{X})=1-\alpha\,]
                                              \end{array}
                                            \right).
\end{equation*}
\vspace{0.05cm}
%where $\partial \overline{L}(\alpha) $ is the $(1-\alpha)$-level set of $\overline{F}$, i.e., $\partial \overline{L}(\alpha) = \partial\{\textbf{x} \in \mathbb{R}^{d}_+ : \overline{F}_{\textbf{X}}(\textbf{x})  \leq 1 - \alpha\}$.
\end{Definition} \vspace{0.15cm}

\noindent
As for $\partial \underline{L}(\alpha)$, under the \emph{regularity conditions},   $\partial \overline{L}(\alpha)=  \{\textbf{x} \in \mathbb{R}^{d}_{+}: \overline{F}(\textbf{x}) = 1-  \alpha\}$  has  Lebesgue-measure zero in $\mathbb{R}^{d}_{+}$ (e.g., see Property 3 in Tibiletti, 1990\nocite{Tibiletti1990}) and  we make sense of  Definition  \ref{VAR multivariate upper}  using the limit  Feller's procedure (see Equations \eqref{limit procedure}-\eqref{VAR multivariate formula}).\\

%\begin{sloppypar}
\noindent
From now on,  we denote  by $\underline{{\rm VaR}}^1_\alpha(\textbf{X})$, $\ldots$, $\underline{{\rm VaR}}^d_\alpha(\textbf{X})$   the  components of the vector  $\underline{{\rm VaR}}_\alpha(\textbf{X})$ and by $\overline{{\rm VaR}}^1_\alpha(\textbf{X})$, $\ldots$, $\overline{{\rm VaR}}^d_\alpha(\textbf{X})$   the  components of the vector  $\overline{{\rm VaR}}_\alpha(\textbf{X})$.\\

Note that if  $\textbf{X}$ is an  exchangeable random vector,  ${\underline{\mbox{VaR}}_{\alpha}^i(\textbf{X})= \underline{\mbox{VaR}}_{\alpha}^j(\textbf{X})}$ and ${\overline{\mbox{VaR}}_{\alpha}^i(\textbf{X})= \overline{\mbox{VaR}}_{\alpha}^j(\textbf{X})}$ for any $i,j =1, \ldots, d$.  Furthermore,  given a univariate random variable $X$,   $\E[X \, | F_X(X)  = \alpha] = \E[X \, | \overline{F}_X(X)  = 1-\alpha] = \rm{VaR}_\alpha(X),$ for all $\alpha$ in $(0,1)$. Hence,  lower-orthant VaR and upper-orthant VaR are the same for (univariate) random variables and Definitions \ref{VAR multivariate} and  \ref{VAR multivariate upper}  can be viewed as  natural multivariate versions of the univariate case. As remarked above,  in Definitions  \ref{VAR multivariate}-\ref{VAR multivariate upper}  instead of considering the whole hyperspace $\partial \underline{L}(\alpha)$  (or  $\partial \overline{L}(\alpha)$), we only focus on the particular point in $\mathbb{R}_+^d$  that matches  the conditional expectation of $\textbf{X}$ given that $\textbf{X}$ falls in $\partial \underline{L}(\alpha)$ (or in $\partial \overline{L}(\alpha)$).\\

%\end{sloppypar}

%\textit{Property of level curves $\partial \overline{L}(\alpha)$  and $\partial \underline{L}(\alpha)$:}
%Let assume the $X_i$ is uniformly distributed in $[0,1]$, for each $i =1, \ldots, d$. One can prove the following relation between $\partial \overline{L}_{\textbf{X}}(\alpha) =  \{\textbf{x} \in \mathbb{R}^{d}_{+}: \overline{F}_{\textbf{X}}(\textbf{x}) = 1-  \alpha\}$ and  $\partial  \underline{L}_{\textbf{X}}(\alpha) =  \{\textbf{x} \in \mathbb{R}^{d}_{+}: F_{\textbf{X}}(\textbf{x}) = \alpha\}$ :
%
%\begin{equation}\label{relatione tra le level curves}
%  \partial \overline{L}_{\textbf{X}}(\alpha)= \mbox{Symmetric}_{\left\{\frac{1}{2}, \ldots, \frac{1}{2}\right\}}  \partial \underline{L}_{1-\textbf{X}}(1-\alpha)
%\end{equation}
%Equation \eqref{relatione tra le level curves}  provides the   relation between the level-curves $\partial \underline{L}$ and $\partial \overline{L}$ using a symmetry in $[0, 1]^d$   with respect to the point $\left\{\frac{1}{2}, \ldots, \frac{1}{2}\right\}$. This geometric result is linked to many properties of   $\underline{{\rm VaR}}_\alpha(\textbf{X})$ and  $\overline{{\rm VaR}}_\alpha(\textbf{X})$ derived in this paper (see, for instance, Equation \eqref{VAR upper usando Cbar}).

\subsection{Invariance properties}  \label{Invariance properties VAR}

In the present section, the aim is to analyze the lower-orthant VaR and upper-orthant VaR  introduced  in Definitions  \ref{VAR multivariate}-\ref{VAR multivariate upper} in terms of classical invariance  properties of risk measures (we refer the interested reader to Artzner \textit{et al.}\nocite{Artzner}, 1999). As these measures are not the same in general for dimension greater or equal to $2$, we also provide some connections between these two measures.\\

%We first derive the following lemma which shows how the level set of multivariate distribution functions and multivariate survival functions are impacted when the underlying random vectors are modified by some increasing/decreasing transformations.

% \begin{Lemma}\label{Transfo_Level_Set}
%Let $\textbf{X} = (X_1, \ldots, X_d)$ be a random vector satisfying  the  regularity conditions and let the function $h$ be such that $h(x_1, \ldots, x_d) = (h_1(x_1), \ldots, h_d(x_d))$.
%\begin{itemize}
%  \item[-] If   $h_1, \ldots, h_d$ are    non-decreasing functions, then the following relations hold
%  \begin{align*}
%\partial \underline{L}_{h(\textbf{X})}(\alpha) &= h\left( \partial \underline{L}_{\textbf{X}}(\alpha) \right), \\
%\partial \overline{L}_{h(\textbf{X})}(\alpha) &= h\left( \partial \overline{L}_{\textbf{X}}(\alpha) \right).
%\end{align*}
%  \item[-]
%If   $h_1, \ldots, h_d$ are    non-increasing functions, then the following relations hold
%  \begin{align*}
%\partial \underline{L}_{h(\textbf{X})}(\alpha) &= h\left( \partial \overline{L}_{\textbf{X}}(1-\alpha) \right), \\
%\partial \overline{L}_{h(\textbf{X})}(\alpha) &= h\left( \partial \underline{L}_{\textbf{X}}(1-\alpha) \right).
%\end{align*}
%\end{itemize}
%\end{Lemma}

We  now introduce the following results (Proposition \ref{passaggio tra le VAR upper et lower con h} and Corollary \ref{passaggio tra le VAR}) that will be useful in order prove  invariance  properties of our risk measures.

 \begin{Proposition}\label{passaggio tra le VAR upper et lower con h}
Let the function $h$ be such that $h(x_1, \ldots, x_d) = (h_1(x_1), \ldots, h_d(x_d))$.
\begin{itemize}
  \item[-] If   $h_1, \ldots, h_d$ are    non-decreasing functions, then the following relations hold
$$
\underline{{\rm VaR}}^{i}_\alpha(h(\textbf{X})) = \E[\,h_i(X_i)\,|\,F_{\textbf{X}}(\textbf{X})= \alpha\,],\;\; i =1, \ldots, d.
$$
\item[-]
If   $h_1, \ldots, h_d$ are    non-increasing functions, then the following relations hold
$$
\underline{{\rm VaR}}^{i}_\alpha(h(\textbf{X})) = \E[\,h_i(X_i)\,|\,\overline{F}_{\textbf{X}}(\textbf{X})= \alpha\,],\;\; i =1, \ldots, d.
$$
\end{itemize}
\end{Proposition}
\vspace{0.1cm}

\textit{Proof: }
From Definition  \ref{VAR multivariate}, $\underline{{\rm VaR}}^{i}_\alpha(h(\textbf{X})) = \E[\,h_i(X_i)\,|\, {F}_{h(\textbf{X})}(h(\textbf{X}))= \alpha\,]$,  for $i =1, \ldots, d$. Since $$F_{h(\textbf{X})}(y_1, \ldots, y_d)=  \left\{
                                               \begin{array}{ll}
                                                F_{\textbf{X}}(h^{-1}(y_1), \ldots, h^{-1}(y_d)), & \mbox{ if } h_1, \ldots, h_d  \mbox{  are    non-decreasing functions,} \\
                                                 \overline{F}_{\textbf{X}}(h^{-1}(y_1), \ldots, h^{-1}(y_d)), & \mbox{ if } h_1, \ldots, h_d  \mbox{  are    non-increasing functions,}
                                               \end{array}
                                             \right.$$
then we obtain the result. $\Box$\vspace{0.17cm}

From Proposition \ref{passaggio tra le VAR upper et lower con h}  one can trivially obtain   the following property which links the multivariate upper-orthant Value-at-Risk and lower-orthant  one.

\begin{Corollary}\label{passaggio tra le VAR}
Let  $h$ be a linear function  such that $h(x_1, \ldots, x_d) = (h_1(x_1), \ldots, h_d(x_d))$.
\begin{itemize}
  \item[-] If   $h_1, \ldots, h_d$ are    non-decreasing functions then then it holds that
  \begin{center}
  $\underline{{\rm VaR}}_\alpha(h(\textbf{X})) = h(\underline{{\rm VaR}}_\alpha(\textbf{X})) \quad $ and $\quad  \overline{{\rm VaR}}_\alpha(h(\textbf{X})) = h(\overline{{\rm VaR}}_\alpha(\textbf{X}))$.
  \end{center}
  \item[-] If   $h_1, \ldots, h_d$ are    non-increasing functions then it holds that
\begin{center}
$\underline{{\rm VaR}}_\alpha(h(\textbf{X})) =  h(\overline{{\rm VaR}}_{1-\alpha}(\textbf{X})) \quad $ and $\quad  \overline{{\rm VaR}}_\alpha(h(\textbf{X})) =  h(\underline{{\rm VaR}}_{1-\alpha}(\textbf{X}))$.
\end{center}
\end{itemize}
\end{Corollary}

 \begin{Example} \label{Link_LO_UO_VaR}
 If $X = (X_1, \ldots, X_d)$ is a random vector with uniform margins and if, for all $i=1,\ldots, d$, we consider the functions $h_i$ such that $h_i(x) = 1-x$, $x\in [0,1]$, then from Corollary \ref{passaggio tra le VAR},
 \begin{equation}
 \label{Eq:Link_VaR_UO_LU}
 \overline{{\rm VaR}}^i_\alpha(\textbf{X}) =  1 -\underline{{\rm VaR}}^i_{1-\alpha}(\textbf{1}-\textbf{X})
  \end{equation}
 for all $i=1, \ldots, d$, where $\textbf{1}-\textbf{X} = (1-X_1, \ldots, 1-X_d)$. In this case, $\overline{{\rm VaR}}_\alpha(\textbf{X})$ is the point reflection of  $\, \underline{{\rm VaR}}_{1-\alpha}({\rm\textbf{1}}-\textbf{X})$  with respect to point $\mathcal{I}$ with coordinates $(\frac{1}{2}, \ldots, \frac{1}{2})$. If $\textbf{X}$ and $\textbf{1}-\textbf{X}$ have the same distribution function, then $X$ is invariant in law by central symmetry and additionally the copula of $X$ and its associated survival copula are the same. In that case
   $\overline{{\rm VaR}}_\alpha(\textbf{X})$ is the point reflection of  $\underline{{\rm VaR}}_{1-\alpha}(\textbf{X})$ with respect to $\mathcal{I}$. This property holds for instance for elliptical copulas or for the Frank copula.
 \end{Example}

Finally, we can state the  following result that  proves positive homogeneity and translation invariance for  our measures.

\begin{Proposition}\label{invarianza var biv}
Consider a random vector $\textbf{X}$ satisfying  the  regularity conditions. For $\alpha \in (0,1)$, the  multivariate upper-orthant and lower-orthant Value-at-Risk  satisfiy the  following properties: \vspace{0.17cm}

\noindent
 Positive Homogeneity:  $\quad  \forall \,\,\,  \textbf{c} \in    \Real^d_+,$
\begin{equation*}
\underline{{\rm VaR}}_\alpha(\textbf{c}   \textbf{X})   =   \textbf{c}    \underline{{\rm VaR}}_\alpha(\textbf{X}),  \quad
% \left(
 %                                             \begin{array}{ll}
  %                                              c_1 \E[X_1|F(\textbf{X})= \alpha]\vspace{0.04cm} \\
   %                                                \quad \quad \quad \quad  \vdots  \vspace{0.04cm} \\
    %                                            c_d \E[X_d|F(\textbf{X})= \alpha]
     %                                         \end{array}
      %                                      \right). \quad
\overline{{\rm VaR}}_\alpha(\textbf{c}  \textbf{X})   =   \textbf{c}   \overline{{\rm VaR}}_\alpha(\textbf{X})
%= \left(
%                                              \begin{array}{ll}
%                                                c_1 \E[X_1|\overline{F}(\textbf{X})= 1-\alpha]\vspace{0.04cm} \\
%                                                   \quad \quad \quad \quad  \vdots  \vspace{0.04cm} \\
%                                                c_d \E[X_d|\overline{F}(\textbf{X})= 1- \alpha]
%                                              \end{array}
%                                            \right).
\end{equation*}
 \noindent
  Translation Invariance: $\quad  \forall \, \,\,  \textbf{c} \in  \Real^d_+,$
 \begin{equation*}
\underline{{\rm VaR}}_\alpha(\textbf{c} + \textbf{X})   =   \textbf{c} + \underline{{\rm VaR}}_\alpha(\textbf{X}), \quad
%= \left(
%                                              \begin{array}{ll}
%                                                c_1 + \E[\,X_1\,|\,F(\textbf{X})= \alpha\,]\vspace{0.04cm} \\
%                                                  \quad \quad \quad \quad  \vdots  \vspace{0.04cm} \\
%                                                c_d + \E[\,X_d\,|\,F(\textbf{X})= \alpha\,]
%                                              \end{array}
%                                            \right).
\overline{{\rm VaR}}_\alpha(\textbf{c} + \textbf{X})   =   \textbf{c} + \overline{{\rm VaR}}_\alpha(\textbf{X})
                                            \end{equation*}
\end{Proposition}
\vspace{0.05cm}

\noindent
 The proof comes down  from Corollary  \ref{passaggio tra le VAR}.

\subsection{Archimedean copula case}\label{Archimedean copula section}

Surprisingly enough, the \underline{VaR} and  $\overline{\VAR}$ introduced in Definitions \ref{VAR multivariate}-\ref{VAR multivariate upper}  can be computed analytically for any $d-$dimensional random vector with an Archimedean copula dependence structure. This is due to McNeil and Ne\v{s}lehov\'{a}'s stochastic representation of Archimedean copulas.

\begin{Proposition} (McNeil and Ne\v{s}lehov\'{a}, 2009)
\label{McNeil_Neslehova}
Let $\textbf{U}= (U_1, \ldots, U_d)$ be distributed according to a $d$-dimensional Archimedian copula with generator $\phi$, then
\begin{equation}\label{Eq_Representation_Archi_1}
\left(\phi(U_1), \ldots, \phi(U_d)\right) \eqd R\textbf{S}\,,
\end{equation}
where $\textbf{S}=(S_1, \ldots, S_d)$ is uniformly distributed on the unit simplex $\left\{\textbf{x}\geq 0 \mid \sum_{k=1}^d x_k = 1\right\}$ and $R$ is an independent non-negative scalar random variable which can be interpreted as the radial part of $\left(\phi(U_1), \ldots, \phi(U_d)\right)$ since $\sum_{k=1}^d S_k = 1$. The random vector $\textbf{S}$ follows a symmetric Dirichlet distribution  whereas the distribution of $R \eqd \sum_{k=1}^d \phi(U_k)$ is directly related to the generator $\phi$ through the inverse Williamson transform of $\phi^{-1}.$\end{Proposition}

Recall that a $d$-dimensional Archimedean copula with generator $\phi$ is defined by $C(u_1, \ldots, u_d) = \phi^{-1}(\phi(u_1)+ \cdots+ \phi(u_d))$, for all $(u_1, \ldots, u_d) \in [0,1]^d.$ Then, the radial part $R$ of representation (\ref{Eq_Representation_Archi_1}) is directly related to the generator $\phi$ and the probability integral transformation of $\textbf{U}$, that is,
\begin{equation*}
R \eqd  \phi(C(\textbf{U})).
\end{equation*}

As a result, any random vector $\textbf{U}=\left(U_1, \ldots, U_d\right)$ which follows an Archimedean copula with generator $\phi$ can be represented as a deterministic function of $C(\textbf{U})$ and an independent  random vector $\textbf{S}=(S_1, \ldots, S_d)$ uniformly distributed on the unit simplex, i.e.,

\begin{equation}\label{Eq_Representation_Archi_2}
\left(U_1, \ldots, U_d\right) \eqd \left(\phi^{-1}\left(S_1\phi\left(C(\textbf{U})\right)\right), \ldots,   \phi^{-1}\left(S_d\phi\left(C(\textbf{U})\right)\right) \right).
\end{equation}

The previous relation allows us to obtain an easily  tractable expression of $\underline{{\rm VaR}}_\alpha(\textbf{X})$ for any random vector $\textbf{X}$ with an Archimedean copula dependence structure.

\begin{Corollary}\label{VaR_Archimedean}
Let $\textbf{X}$ be a $d$-dimensional random vector with marginal distributions $F_1, \ldots, F_d$. Assume that the dependence structure of $\textbf{X}$ is given by an Archimedian copula with generator $\phi$. Then, for any $i=1,\ldots, d,$
\begin{equation}\label{Expression_VaR_Arch}
\underline{{\rm VaR}}_\alpha^i(\textbf{X}) = \E\left[F_i^{-1}\left(\phi^{-1}(S_i \phi(\alpha))\right)\right]
\end{equation}
where $S_i$ is a random variable with ${\rm Beta}(1, d-1)$ distribution.
\end{Corollary}

\noindent
\textit{Proof: }
 Note that  $\textbf{X}$ is distributed as $(F_1^{-1}(U_1), \ldots, F_d^{-1}(U_d))$ where $\textbf{U} = (U_1, \ldots, U_d)$ follows an Archimedean copula $C$ with generator $\phi$. Then, each component  $i=1,\ldots, d\,$ of the multivariate risk measure introduced in Definition \ref{VAR multivariate} can be expressed as  $\underline{{\rm VaR}}_\alpha^i(\textbf{X}) = \E\left[F_i^{-1}(U_i) \mid C(\textbf{U})=\alpha \right]$.
 Moreover, from representation (\ref{Eq_Representation_Archi_2})  the following relation holds
\begin{equation}\label{Archi_Cond_Distr}
 \left[\textbf{U}\mid C(\textbf{U}) = \alpha\right] \eqd \left(\phi^{-1}\left(S_1\phi\left(\alpha\right)\right), \ldots,   \phi^{-1}\left(S_d\phi\left(\alpha\right)\right) \right)
 \end{equation}
 since  $\textbf{S}$ and $C(\textbf{U})$ are stochastically independent. The result comes down from the fact that the random vector $(S_1, \ldots, S_d)$ follows a symmetric Dirichlet distribution.
 $\Box$\\

Note that, using (\ref{Archi_Cond_Distr}), the marginal distributions of $\textbf{U}$ given $C(\textbf{U})=\alpha$ can be expressed in a very simple way, that is,  for any $k=1,\ldots, d,$
\begin{equation}
\label{Marginal_Cond_Distr_Archi}
\Pbb(U_k\leq u \mid C(\textbf{U})=\alpha) = \left(1 - \frac{\phi(u)}{\phi(\alpha)}\right)^{d-1} \,\, \quad \mbox{ for } \,\, 0<\alpha<u<1.
 \end{equation}
The latter relation derives from the fact that $S_k,$ which is  ${\rm Beta}(1, d-1)$- distributed, is such that $S_k \eqd 1-V^{\frac{1}{d-1}}$ where $V$ is uniformly-distributed on $(0,1)$. \\

We now  adapt  Corollary \ref{VaR_Archimedean}  for the multivariate upper-orthant Value-at-Risk, i.e.,  $\overline{{\rm VaR}}_\alpha$.

\begin{Corollary}\label{VaR_Archimedean upper}
Let $\textbf{X}$ be a $d$-dimensional random vector with marginal survival distributions $\overline{F}_1, \ldots, \overline{F}_d$.  Assume that  the survival copula of $\textbf{X}$ is an Archimedean   copula with  generator $\phi$.  Then, for any $i=1,\ldots, d,$
\begin{equation}\label{Expression_VaR_Arch upper}
\overline{{\rm VaR}}_\alpha^i(\textbf{X}) = \E\left[\overline{F}_{i}^{-1}\left(\phi^{-1}(S_i \phi(1-\alpha))\right)\right]
\end{equation}
where $S_i$ is a random variable with ${\rm Beta}(1, d-1)$ distribution.
\end{Corollary}

\noindent
\textit{Proof: }
 Note that  $\textbf{X}$ is distributed as $(\overline{F}_1^{-1}(U_1), \ldots, \overline{F}_d^{-1}(U_d))$ where $\textbf{U} = (U_1, \ldots, U_d)$ follows an Archimedean copula $\overline{C}$ with generator $\phi$. Then, each component  $i=1,\ldots, d\,$ of the multivariate risk measure introduced in Definition  \ref{VAR multivariate upper} can be expressed as  $\overline{{\rm VaR}}_\alpha^i(\textbf{X}) = \E\left[\overline{F}_i^{-1}(U_i) \mid \overline{C}(\textbf{U})=1-\alpha \right]$.
%Let ${U}_i= \overline{F}_{X_{i}}(X_i)$, for  $i =1, \ldots, d$. Using Definition \ref{VAR multivariate upper}, one can write:
%\begin{equation}\label{VAR upper usando Cbar importante}
%\overline{{\rm VaR}}^i_\alpha(\textbf{X})=\E[\overline{F}^{-1}_{i}({U}_k)|\overline{C}({U}_1, \ldots, \overline{U}_d)= 1-\alpha]% = \E[F^{-1}_{i}(1-U_i) |\overline{C}_{\textbf{X}}(U_1, \ldots, U_d)= 1-\alpha].
%\end{equation}
%where $\mathbf{{U}} := ({U}_1, \ldots,{U}_d)$ admits the Archimedean copula $\overline{C}$ as multivariate distribution function.
Then, relation \ref{Archi_Cond_Distr} also holds for $\mathbf{{U}}$ and $\overline{C}$, i.e.,
%Using Equation \eqref{VAR upper usando Cbar importante} we have  $\overline{{\rm VaR}}^k_\alpha(\textbf{X})=\E[\overline{F}^{-1}_{k}(U_k) \,|\,\overline{C}_{\textbf{X}}(U_1, \ldots, U_d)= 1-\alpha]$.
%Since representation in \eqref{Eq_Representation_Archi_2}  holds for the  Archimedean co\-pu\-la  $\overline{C}$, this implies that
  $\left[\textbf{U}\mid \overline{C}(\textbf{U}) = 1-\alpha\right] \eqd \left(\phi^{-1}\left(S_1\phi\left(1-\alpha\right)\right), \ldots,   \phi^{-1}\left(S_d\phi\left(1-\alpha\right)\right) \right)$. Hence the result.~$\Box$\\

%\begin{Example}\label{marges uniforms}
%\rm{
%If we assume that $X_i$ is uniformly distributed in $[0,1]$, for each $i =1, \ldots, d$, from \eqref{VAR upper usando Cbar importante}  we obtain
%\begin{equation} \label{VAR upper usando Cbar}
%  \overline{{\rm VaR}}^i_\alpha(\textbf{X})= 1- \E[ U_i \,|\,\overline{C}_{\textbf{X}}(U_1, \ldots, U_d)= 1-\alpha]= 1- \underline{{\rm VaR}}^i_{1 -\alpha}(1-\textbf{X}).
%  \end{equation}
%  Remark that \eqref{VAR upper usando Cbar} directly comes down from Corollary \ref{passaggio tra le VAR} using the non-increasing transformation functions $U_i= 1- X_i$, for $i =1, \ldots, d$. }
%\end{Example}
%
In the following, from \eqref{Expression_VaR_Arch} and \eqref{Expression_VaR_Arch upper},  we derive analytical expressions of the  lower-orthant and the upper-orthant  Value-at-Risk for a random vector  $\mathbf{X}=(X_1, \ldots, X_d)$ distributed as a particular Archimedean copula.  Let us first remark that , as Archimedean copulas are exchangeable, the components of $\underline{\rm{VaR}}$ (resp. $\overline{\rm{VaR}}$) are the same. Moreover, as far as closed-form expressions are available for the lower-orthant $\underline{\rm{VaR}}$ of $\mathbf{X}$, it is also possible to derive an analogue expression for the upper-orthant $\overline{\rm{VaR}}$ of $\mathbf{\tilde{X}} = (1-X_1, \ldots, 1-X_d)$ since  from Example \ref{Link_LO_UO_VaR}
\begin{equation} \label{Link_Arch_Copula}
 \overline{{\rm VaR}}^i_{\alpha}(\mathbf{\tilde{X}}) = 1 - \underline{{\rm VaR}}^i_{1-\alpha}(\mathbf{X}).
\end{equation}
%of a bivariate vector $(\tilde{X},\tilde{Y})$ with uniform margins and which admits a Clayton copula as survival copula. Indeed, thanks to Corollary \ref{VaR_Archimedean upper} or Example \ref{Link_LO_UO_VaR},
%\begin{equation} \label{Link_Arch_Copula}
% \overline{{\rm VaR}}^1_{\alpha, \theta}(\tilde{X}, \tilde{Y}) = 1 - \underline{{\rm VaR}}^1_{1-\alpha, \theta}(X, Y)
%\end{equation}
% where $(X,Y) := (1-\tilde{X}, 1-\tilde{X})$ is distributed as the same Clayton copula with parameter $\theta$.% with dependence parameter $\theta$.
% \\

\underline{Clayton family in dimension $2$:}\\

 As a matter of example, let us now consider the Clayton family of  bivariate copulas. This family is interesting since it contains the counter-monotonic, the independence and the comonotonic copulas as particular cases.
Let $(X,Y)$ be a random vector distributed as a Clayton copula with parameter $\theta \geq -1$. Then, $X$ and $Y$ are uniformly-distributed on $(0,1)$  and the joint distribution function $C_\theta$ of $(X,Y)$ is such that

\begin{equation}\label{clayton}
C_\theta(x,y)= (\max\{x^{-\theta}+y^{-\theta}-1, 0\})^{-\frac{1}{\theta}}, \quad \mbox{ for } \theta  \geq -1, \,\,\,   \,\, (x,y) \in [0,1]^2.
\end{equation}

\noindent
%Let us stress that since  $X$ and $Y$ are exchangeable, the two components of the multivariate $\underline{\rm{VaR}}$ and $\overline{\rm{VaR}}$ are identical.
Table  \ref{VAR different copula}  gives analytical expressions for the first (equal to the second) component of  $\underline{\rm{VaR}}$
 %(third column) and for the first (equal to the second) component of $\overline{\rm{VaR}}$  (fourth column)
 as a function of the risk level $\alpha$ and the dependence parameter $\theta$.
 For $\theta = -1$ and $\theta = \infty $ we obtain the Fr\'echet-Hoeffding lower and upper bounds: $W(x,y)=  \max\{x +y -1, 0\}$ (counter-monotonic copula) and $M(x,y)= \min\{x, y\}$ (comonotonic random copula) respectively. The settings $\theta=0$ and $\theta=1$ correspond to degenerate cases. For $\theta = 0$ we have the independence copula $\Pi(x,y)=x\,y$. For $\theta=1$, we obtain the copula denoted by $\frac{ \Pi}{\Sigma - \Pi}$ in  Nelsen (1999\nocite{NelsenLibro}),  where $\frac{ \Pi}{\Sigma - \Pi}(x,y)= \frac{x \,y}{x + y - x \,y}$.

\begin{table}[!ht]
 \begin{minipage}[h!]{0.99\linewidth}\centering
\begin{tabular}{| c | c| c | c | c |c |}
 \hline \centering
     {\Large\strut}  Copula   &  $\theta$ &   $\underline{{\rm VaR}}^1_{\alpha, \theta}(X,Y)$\\  \hline %\vspace{0.02cm}
       {\Large\strut}   Clayton  $C_\theta$ & $(-1, \infty)$ &  $\frac{\theta}{\theta-1}{\frac {{\alpha}^{\theta} -\alpha}{  {\alpha}^{\theta} -1 }}$ \\
\vspace{0.02cm}
         {\large\strut}   Counter-monotonic $W$ &   $-1$ & $\frac{1+ \alpha}{2}$    \\
         \vspace{0.02cm}
       {\large\strut}    Independent $\Pi$   & $0$ &   ${\frac {\alpha-1}{\ln   \alpha  }}$  \\
     {\large\strut}  $\frac{ \Pi}{\Sigma - \Pi} $ & $1$ &   ${\frac {\alpha\,\ln  \alpha  }{\alpha-1}}$  \\
       \vspace{0.02cm}
     {\large\strut}     Comonotonic  $M$ & $\infty$ & $ \alpha$ \\ %\vspace{0.02cm}
\hline
   \end{tabular}
    \caption{{\small $\underline{{\rm VaR}}^1_{\alpha, \theta}(X,Y)$ for different dependence structures. \vspace{0.2cm}}}
   \label{VAR different copula}
      \end{minipage}
\end{table}

%\begin{table}[!ht]
% \begin{minipage}[h!]{0.99\linewidth}\centering
%\begin{tabular}{| c | c| c | c | c |c |c |}
% \hline \centering
%     {\Large\strut}  Copula   &  $\theta$ &   $\underline{{\rm VaR}}^1_{\alpha, \theta}(X,Y)$ &   $\overline{{\rm VaR}}^1_{\alpha, \theta}(X,Y)$ \\  \hline %\vspace{0.02cm}
%       {\huge\strut}   Clayton  $C_\theta$ & $(-1, \infty)$ &  $\frac{\theta}{\theta-1}{\frac {{\alpha}^{\theta} -\alpha}{  {\alpha}^{\theta} -1 }}$&
%        $-\frac{(\theta \alpha -1)+(1- \alpha)^\theta}{(\theta-1) ((1-\alpha)^\theta -1)} $     \\
%\vspace{0.02cm}
%         {\Large\strut}   Counter-monotonic $W$ &   $-1$ & $\frac{1+ \alpha}{2}$  & $\frac{\alpha}{2}$  \\
%         \vspace{0.02cm}
%       {\Large\strut}    Independent $\Pi$   & $0$ &   ${\frac {\alpha-1}{\ln   \alpha  }}$ & $1+ {\frac {\alpha}{\ln(1-\alpha)}}$ \\
%     {\Large\strut}  $\frac{ \Pi}{\Sigma - \Pi} $ & $1$ &   ${\frac {\alpha\,\ln  \alpha  }{\alpha-1}}$ & $1 + \frac{\ln(1-\alpha) (1-\alpha)}{\alpha} $  \\
%       \vspace{0.02cm}
%     {\Large\strut}     Comonotonic  $M$ & $\infty$ & $ \alpha$  &$\alpha$  \\ %\vspace{0.02cm}
%\hline
%   \end{tabular}
%    \caption{{\small $\underline{{\rm VaR}}^1_{\alpha, \theta}(X,Y)$ and $\overline{{\rm VaR}}^1_{\alpha, \theta}(X,Y)$ for different dependence structures.}}
%   \label{VAR different copula}
%      \end{minipage}
%\end{table}

\begin{figure}[h!]
\begin{minipage}[b]{0.5\linewidth}\centering
\includegraphics[width=8.1cm]{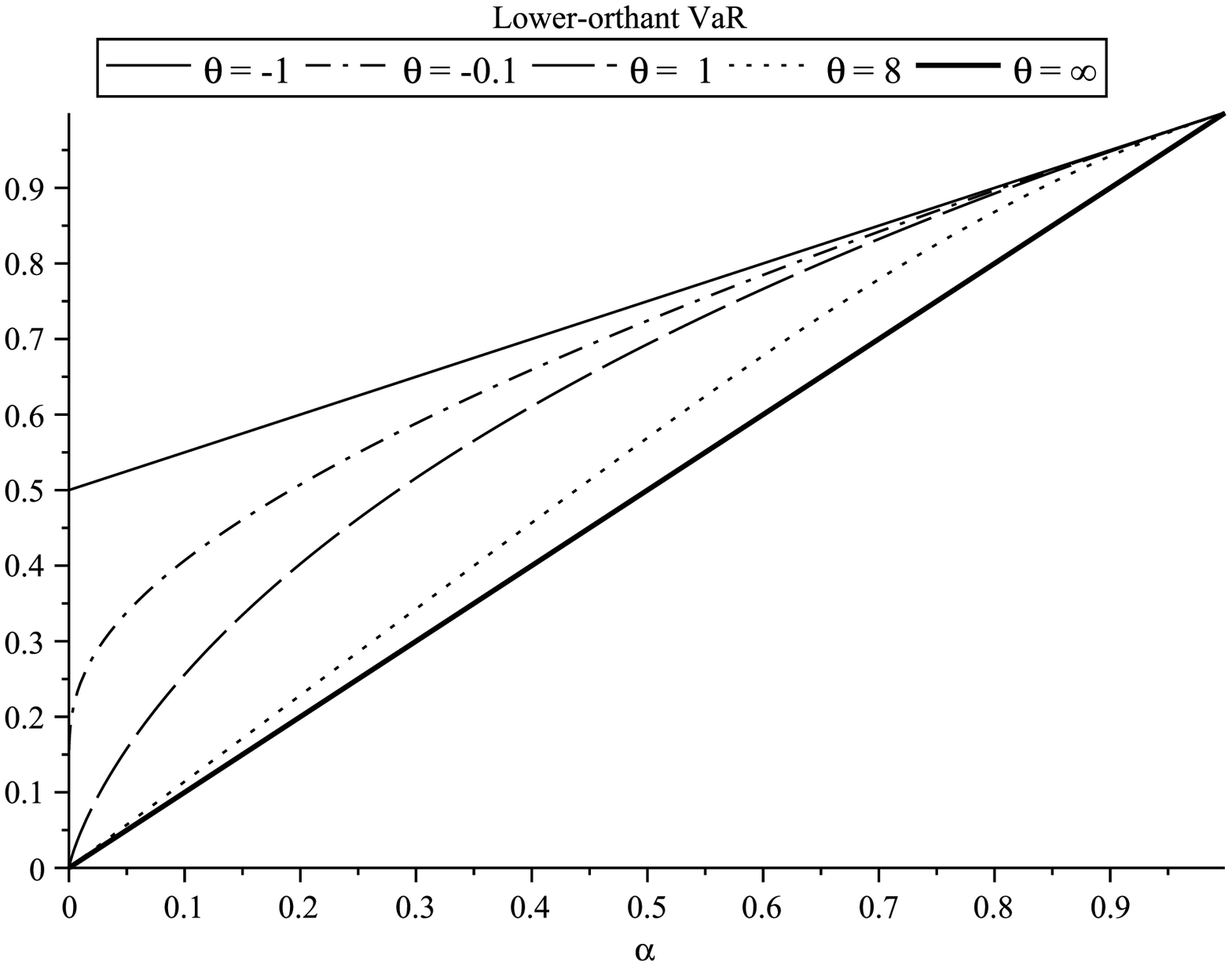}
\end{minipage}
\begin{minipage}[b]{0.4\linewidth}\centering
\includegraphics[width=8cm]{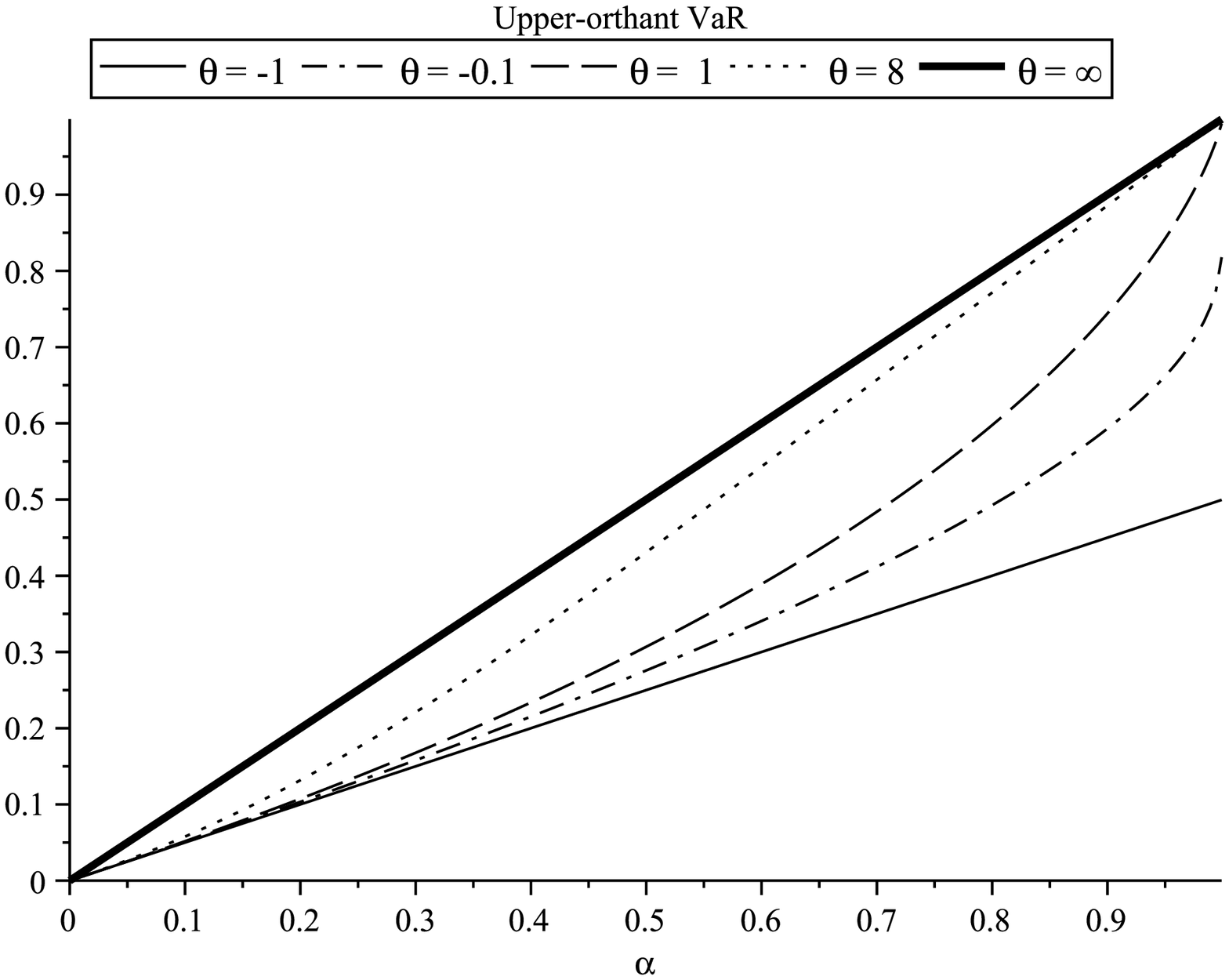}
\end{minipage}
 \caption{{ \small Behavior of  $\underline{{\rm VaR}}^1_{\alpha, \theta}(X,Y)= \underline{{\rm VaR}}^2_{\alpha, \theta}(X,Y)$ (left) and $\overline{{\rm VaR}}^1_{\alpha, \theta}(1-X,1-Y)= \overline{{\rm VaR}}^2_{\alpha, \theta}(1-X,1-Y)$ (right)  with respect to risk level $\alpha$ for different values of  dependence parameter $\theta$. The random vector $(X,Y)$ follows a Clayton copula distribution with parameter $\theta$.  Note that, due to Equation \ref{Link_Arch_Copula}, the two graphs are symmetric with respect to the point $(\frac{1}{2},\frac{1}{2})$}}
\label{VAR bivariate Clayton grafico}
\end{figure}

 Interestingly, one can readily show that  $\frac{\partial \underline{{\rm VaR}}^1_{\alpha, \theta}}{\partial \alpha} \geq  0$ and $\frac{\partial \underline{{\rm VaR}}^1_{\alpha, \theta}}{\partial \theta} \leq 0$, for   $\theta  \geq -1$ and $\alpha \in (0,1)$. This proves that, for Clayton-distributed random couples, the components of our multivariate \underline{VaR} are increasing functions of the risk level $\alpha$ and decreasing functions of the dependence parameter $\theta$. Note also that the multivariate \underline{VaR} in the comonotonic case corresponds to the vector composed of the univariate VaR associated with each component.   These properties are illustrated in Figure \ref{VAR bivariate Clayton grafico} (left)  where $\underline{{\rm VaR}}^1_{\alpha, \theta}(X,Y)$ is plotted as a function of the risk level $\alpha$ for different values of the parameter $\theta$. Observe that an increase of the dependence parameter $\theta$ tends to lower the \underline{VaR} up to the perfect dependence case where $\underline{{\rm VaR}}_{\alpha, \theta}^1(X,Y)= \underline{{\rm VaR}}_{\alpha}(X)=\alpha$. The latter empirical behaviors will be formally confirmed in next sections. \\

In the same framework, using Equation  \ref{Link_Arch_Copula}, one can readily show that  $\frac{\partial \overline{{\rm VaR}}^1_{\alpha, \theta}}{\partial \alpha}  \geq  0$ and $\frac{\partial \overline{{\rm VaR}}^1_{\alpha, \theta}}{\partial \theta} \geq 0$, for   $\theta  \geq -1$ and $\alpha \in (0,1)$. This proves that, for random couples with uniform margins and Clayton survival copula, the components of our multivariate $\overline{\VAR}$ are increasing functions both of the risk level $\alpha$ and of the dependence parameter $\theta$.   Note also that the multivariate  $\overline{\VAR}$ in the comonotonic case corresponds to the vector composed of the univariate VaR associated with each component.    These properties are illustrated in Figure \ref{VAR bivariate Clayton grafico} (right)  where $\overline{{\rm VaR}}^1_{\alpha, \theta}(X,Y)$ is plotted as a function of the risk level $\alpha$ for different values of the parameter $\theta$. Observe that, contrary to the lower-orthant  $\underline{\rm{VaR}}$, an increase of the dependence parameter $\theta$ tends to increase  the  $\overline{\VAR}$.  The upper bound is represented by the perfect dependence case where $\overline{{\rm VaR}}_{\alpha, \theta}^1(X,Y)= {\rm VaR}_{\alpha}(X)=\alpha$. The latter empirical behaviors will be formally confirmed in next sections.\\

\underline{Ali-Mikhail-Haq in dimension $2$:}\\

Let $(X,Y)$ be a random vector distributed as a Ali-Mikhail-Haq copula with parameter $\theta \in [-1, 1)$. In particular, the marginal distribution of $X$ and $Y$ are uniform. Then, the distribution function $C_\theta$ of $(X,Y)$ is such that
\begin{equation*}\label{Ali}
C_\theta(x,y)=  \frac{x\,y}{1- \theta\,(1-x)(1-y)}, \quad \mbox{ for } \theta  \in [-1, 1), \,\,\,   \,\, (x,y) \in [0,1]^2.
\end{equation*}
Using Corollary \ref{VaR_Archimedean}, we give in Table  \ref{VAR different copula Ali} analytical expressions for the first (equal to the second) component of the $\underline{\VAR}$,   i.e., $\underline{{\rm VaR}}^1_{\alpha, \theta}(X,Y)$.   When $\theta = 0$ we obtain the independence copula $\Pi(x,y)=x\,y$.

   \vspace{0.1cm}
\begin{table}[!ht]
 \begin{minipage}[h!]{0.99\linewidth}\centering
\begin{tabular}{| c | c| c | c | c |c |}
 \hline \centering
     {\Large\strut}  Copula   &  $\theta$ &   $\underline{{\rm VaR}}^1_{\alpha, \theta}(X,Y)$\\  \hline %\vspace{0.02cm}
       {\Huge\strut}  Ali-Mikhail-Haq copula  $C_\theta$ & $[-1, 1)$ &  ${\frac {\left( \theta -1
 \right) \ln  \left( 1-\theta(1-\alpha) \right)   }{ \theta \left( \ln  \left( 1-\theta(1-
\alpha) \right)  -\ln  \left( \alpha \right)  \right)}}
$ \\
\vspace{0.02cm}
              {\Large\strut}    Independent $\Pi$   & $0$ &   ${\frac {\alpha-1}{\ln  \left( \alpha \right) }}$  \\
     \hline
   \end{tabular}\vspace{0.1cm}
    \caption{{\small $\underline{{\rm VaR}}^1_{\alpha, \theta}(X,Y)$ for a bivariate Ali-Mikhail-Haq copula. \vspace{0.2cm}}}
   \label{VAR different copula Ali}
      \end{minipage}
\end{table}

%\begin{table}[!ht]
% \begin{minipage}[h!]{0.99\linewidth}\centering
%\begin{tabular}{| c | c| c | c | c |c |c|}
% \hline \centering
%     {\Large\strut}  Copula   &  $\theta$ &   $\underline{{\rm VaR}}^1_{\alpha, \theta}(X,Y)$ & $\overline{{\rm VaR}}^1_{\alpha, \theta}(X,Y)$ \\  \hline %\vspace{0.02cm}
%       {\Huge\strut}  Ali-Mikhail-Haq copula  $C_\theta$ & $[-1, 1)$ &  ${\frac {\ln  \left( 1-\theta+\theta\,\alpha \right)  \left( -1+\theta
% \right) }{ \left( -\ln  \left( \alpha \right) +\ln  \left( 1-\theta+
%\theta\,\alpha \right)  \right) \theta}}  $ &  ${\frac {-\theta\,\ln  \left( 1-\alpha \right) +\ln  \left( 1-\theta\,
%\alpha \right) }{ \left( -\ln  \left( 1-\alpha \right) +\ln  \left( 1-
%\theta\,\alpha \right)  \right) \theta}}$ \\
%\vspace{0.02cm}
%              {\Large\strut}    Independent $\Pi$   & $0$ &   ${\frac {\alpha-1}{\ln  \left( \alpha \right) }}$ &  $1+ {\frac {\alpha}{\ln(1-\alpha)}}$ \\
%     \hline
%   \end{tabular}
%    \caption{{\small $\underline{{\rm VaR}}^1_{\alpha, \theta}(X,Y)$ and $\overline{{\rm VaR}}^1_{\alpha, \theta}(X,Y)$ for a bivariate Ali-Mikhail-Haq copula. \vspace{0.2cm}}}
%   \label{VAR different copula Ali}
%      \end{minipage}
%\end{table}
%

\underline{Clayton family in dimension $3$:}\\

We now consider a $3$-dimensional vector $\textbf{X}= (X_1,X_2, X_3)$ with Clayton copula and parameter  $ \theta > - \frac{1}{2}$ (see Remark \ref{theta e dimensione})  and uniform marginals. In this case we give an analytical expression of  $\underline{{\rm VaR}}^i_{\alpha, \theta}(X_1,X_2, X_3)$ for $i=1,2,3$. Results are given in Table \ref{VAR different copula dim 3}.\\

\begin{table}[!ht]
 \begin{minipage}[h!]{0.99\linewidth}\centering
\begin{tabular}{| c | c| c | c | c |c |}
 \hline \centering
     {\Large\strut}  Copula   &  $\theta$ &   $\underline{{\rm VaR}}^i_{\alpha, \theta}(X_1,X_2, X_3)$\\  \hline %\vspace{0.02cm}
       {\huge\strut}   Clayton  $C_\theta$ & $(-1/2, \infty)$ &  $2\,{\frac {\theta\, \left( (\theta-1) {\alpha}^{2\,
\theta} + (1-2\theta) {\alpha}^{\theta} + \theta \alpha \right) }{ \left( 2\,\theta-1 \right)  \left( \theta -1 \right)
 \left( {\alpha}^{2\,\theta} -2\,{\alpha}^{\theta} + 1\right) }}$ \\
\vspace{0.02cm}
                {\huge\strut}    Independent $\Pi$   & $0$ &   $-2\,{\frac {1-\alpha+\ln  \left( \alpha \right) }{ \left( \ln  \left(
\alpha \right)  \right) ^{2}}}$  \\
     %{\huge\strut}  $\frac{ \Pi}{\Sigma - \Pi} $ & $1$ &   $-2\,{\frac {\alpha\, \left( 1-\alpha+\ln  \left( \alpha \right)
 %\right) }{ \left( \alpha-1 \right) ^{2}}}$  \\
      % \vspace{0.02cm}
    % {\large\strut}     Comonotonic  $M$ & $\infty$ & $ \alpha$
  %  \\ %\vspace{0.02cm}
\hline
   \end{tabular}\vspace{0.1cm}
    \caption{{\small $\underline{{\rm VaR}}^1_{\alpha, \theta}(X_1,X_2, X_3)$ for different dependence structures. \vspace{0.2cm}}}
   \label{VAR different copula dim 3}
      \end{minipage}
\end{table}

 As in the bivariate case above,  one can readily show that
 $\frac{\partial \underline{{\rm VaR}}^1_{\alpha, \theta}}{\partial \alpha} \geq  0, $  $\frac{\partial \underline{{\rm VaR}}^1_{\alpha, \theta}}{\partial \theta} \leq 0$ when $\mathbf{X}$ is distributed as a $3-$dimensional  Clayton copula. In addition,  using Equation  \ref{Link_Arch_Copula},
    $\frac{\partial \overline{{\rm VaR}}^1_{\alpha, \theta}}{\partial \alpha} \geq 0$ and $\frac{\partial \overline{{\rm VaR}}^1_{\alpha, \theta}}{\partial \theta}  > 0$  when  $\mathbf{X}$ admits a trivariate Clayton survival copula.
    Then, the results obtained above in the bivariate case are confirmed also in higher dimension. These empirical behaviors will be formally confirmed in next sections. \\

\subsection{Comparison of univariate and multivariate  VaR} \label{proprieta VAR bidimensionale}

Note that, using a change of variable, each component of the multivariate  VaR  can be represented as an integral transformation of the   associated univariate VaR. Let us denote by $F_{X_i}$ the marginal distribution functions of $X_i$ for $i=1, \ldots, d$ and by $C$ (resp. $\overline{C}$) the copula    (resp. the survival copula) associated with $\textbf{X}$. Using the Sklar's theorem we have $F(x_1,\ldots, x_d)=C(F_{X_1}(x_1), \ldots, F_{X_d}(x_d))$ (see Sklar, 1959\nocite{Sklar}). Then the random variables $U_i$ defined by $U_i=F_{X_i}(X_i)$, for $i=1, \ldots, d$,  are uniformly distributed and their joint distribution is equal to $C$. Using these notations and since $F^{-1}_{X_i}(\gamma)= \VAR_\gamma(X_i)$, we get

\begin{equation}\label{VAR XY integrale 1}
\underline{\VAR}_{\alpha}^i(\textbf{X}) = \frac{1}{K'(\alpha)}\int_{\alpha}^1   \VAR_\gamma(X_i)  f_{(U_i,C(\textbf{U}))}(\gamma, \alpha) \, d\gamma,
\end{equation}
\begin{sloppypar}
\noindent
for $i=1, \ldots, d,$ where $K'$ is the density of the Kendall distribution associated with copula $C$ and $f_{(U_i,C(\textbf{U}))}$  is the density function of the bivariate vector $(U_i, C(\textbf{U}))$.  \end{sloppypar}

As for the upper-orthant VaR, let $V_i= \overline{F}_{X_{i}}(X_i)$, for  $i =1, \ldots, d$. %$X_i = \overline{F}^{-1}_{X_{i}}(V_i)$.
Using these notations and since  $\overline{F}^{-1}_{X_{i}}(\gamma) = \VAR_{1-\gamma}(X_i)$, we get
%Since $\overline{{\rm VaR}}^i_\alpha(\textbf{X})=\E[\overline{F}^{-1}_{i}(U_k)|\overline{C}_{\textbf{X}}(U_1, \ldots, U_d)= 1-\alpha]$ for  $i =1, \ldots, d$,    we get
\begin{equation}\label{VAR XY integrale 1 upper}
\overline{\VAR}_{\alpha}^i(\textbf{X}) =
\frac{1}{K'_{\overline{C}}(1-\alpha)}\int_{0}^{1-\alpha} \VAR_{1-\gamma}(X_i)  f_{(V_i,\overline{C}(\textbf{V}))}(\gamma, 1-\alpha) \, d\gamma,
\end{equation}
%\begin{equation}\label{VAR XY integrale 1 upper}
%\overline{\VAR}_{\alpha}^i(\textbf{X}) =
%\frac{1}{K'_{\overline{C}}(1-\alpha)}\int_{0}^{1-\alpha} \overline{F}^{-1}_{X_i}(\gamma)  f_{(V_i,\overline{C}(\textbf{V}))}(\gamma, 1-\alpha) \, d\gamma,
%\end{equation}
\begin{sloppypar}
\noindent
 for $i=1, \ldots, d$, where $K'_{\overline{C}}$ the density of the Kendall distribution associated with the survival copula~$\overline{C}$ and $f_{(V_i,\overline{C}(\textbf{V}))}$ is the density function of the bivariate vector~$(V_i, \overline{C}(\textbf{V}))$. \end{sloppypar}
Remark that the bounds of integration  in  \eqref{VAR XY integrale 1} and \eqref{VAR XY integrale 1 upper} derive from the geometrical properties of the considered level curve, i.e., $\partial \underline{L}(\alpha)$ (resp. $\partial \overline{L}(\alpha)$) is inferiorly (resp. superiorly) bounded by the marginal univariate quantile functions at level $\alpha$.
\vspace{0.1cm}

 The following proposition allows us to compare the multivariate  lower-orthant and upper-orthant \emph{Value-at-Risk} with the corresponding univariate VaR.

\begin{Proposition}\label{VAR XY respect to the univariate VaR}
Consider a random vector $\textbf{X}$ satisfying  the  regularity conditions. Assume that its multivariate distribution function $F$   is quasi concave\footnote{A function  $F$   is quasi concave if  the upper level sets of $F$ are convex sets. Tibiletti (1995\nocite{Tibiletti2}) points out families of distribution functions which satisfy the property of quasi concavity.  For instance, multivariate elliptical distributions and Archimedean copulas are quasi concave functions (see Theorem 4.3.2 in   Nelsen, 1999\nocite{NelsenLibro} for proof in dimension $2$;   Proposition 3 in Tibiletti, 1995\nocite{Tibiletti2},   for proof in dimension $d$).}.   Then,  for all $\, \alpha \in (0,1),$ the following  inequalities hold

\begin{equation}\label{confronto var univ1}
    \overline{{\rm VaR}}^i_{\alpha}(\textbf{X})  \leq {\rm VaR}_\alpha(X_i) \leq \underline{{\rm VaR}}^i_{\alpha}(\textbf{X}),
\end{equation}
 for $i=1, \ldots, d$.
 \end{Proposition}
\vspace{0.1cm}

\noindent
\textit{Proof:}  Let $\alpha \, \in \, (0,1)$.   From the definition of the accumulated probability, it is easy  to show that $\partial \underline{L}(\alpha)$ is inferiorly bounded by the marginal univariate quantile functions.    Moreover,  recall that $\underline{L}(\alpha)$ is a convex set in $\mathbb{R}^d_+$ from the quasi concavity of $F$ (see Section 2 in Tibiletti, 1995\nocite{Tibiletti2}). Then, for all $\textbf{x}=(x_1,\ldots, x_d) \in \partial \underline{L}(\alpha)$, $x_1 \geq \underline{\VAR}_{\alpha}(X_1), \cdots, x_d \geq \underline{\VAR}_{\alpha}(X_d)$ and trivially,  $\underline{{\rm VaR}}^i_{\alpha}(\textbf{X})$   is  greater than ${\rm VaR}_\alpha(X_i)$,  for $i=1, \ldots, d$.  Then  $\underline{{\rm VaR}}^i_{\alpha}(\textbf{X}) \geq{\rm VaR}_\alpha(X_i)$,  for all  $\, \alpha \in (0,1)$ and $i=1, \ldots, d$.
Analogously, from the definition of the survival accumulated probability, it is easy  to show that $\partial \overline{L}(\alpha)$ is superiorly bounded by the marginal univariate quantile functions at level $\alpha$.    Moreover,  recall that $\overline{L}(\alpha)$ is a convex set in $\mathbb{R}^d_+$. Then, for all $\textbf{x}=(x_1,\ldots, x_d) \in \partial \overline{L}(\alpha)$, $x_1 \leq  {\VAR}_{\alpha}(X_1), \cdots, x_d  \leq  {\VAR}_{\alpha}(X_d)$ and trivially,
 $ \overline{{\rm VaR}}^i_{\alpha}(\textbf{X})$   is  smaller than ${\rm VaR}_{\alpha}(X_i)$,  for all  $\, \alpha \in (0,1)$ and $i=1, \ldots, d$.     Hence the result  $\Box$\vspace{0.17cm}

\noindent
Proposition \ref{VAR XY respect to the univariate VaR} states that the multivariate lower-orthant  $\underline{{\rm VaR}}_{\alpha}(\textbf{X})$  (resp. the multivariate upper-orthant  $\overline{{\rm VaR}}_{\alpha}(\textbf{X})$) is more  conservative (resp. less conservative) than the vector composed of  the univariate $\alpha$-\emph{Value-at-Risk} of marginals. Furthermore,   we can prove  that the previous  bounds  in \eqref{confronto var univ1} are  reached for comonotonic random vectors.

\begin{Proposition}\label{comonotonic caseVAR XY respect to the univariate VaR}
Consider a comonotonic non-negative random vector $\textbf{X}$. Then,  for all $\, \alpha \in (0,1),$ it holds that
\begin{equation*}
    \overline{{\rm VaR}}^i_{\alpha}(\textbf{X}) = {\rm VaR}_\alpha(X_i) = \underline{{\rm VaR}}^i_{\alpha}(\textbf{X}),
\end{equation*}
for $i=1, \ldots, d$.
\end{Proposition}
\vspace{0.1cm}
\noindent
\textit{Proof: } Let $\alpha \, \in \, (0,1)$.  If $\textbf{X}=(X_1, \ldots, X_d)$ is a   comonotonic non-negative random vector then there exists a random variable $Z$ and  $d$ increasing functions  $g_1, \ldots, g_d$ such that $\textbf{X}$ is equal to $(g_1(Z), \ldots, g_d(Z))$ in distribution. So the set $\{(x_1,\ldots, x_d): F(x_1,\ldots, x_d)= \alpha\}$ becomes $\{(x_1,\ldots, x_d):  $ $\min\{g^{-1}_1(x_1),  \ldots, g^{-1}_d(x_d)\}$ $= Q_Z(\alpha) \},$ where $Q_Z$ is the quantile function of $Z$.  So, trivially, $\underline{{\rm VaR}}^i_{\alpha}(\textbf{X})=  \E[\,X_i\,|\,F(\textbf{X})= \alpha\,] = Q_{X_i}(\alpha)$,  for $i=1, \ldots, d$ and $\overline{{\rm VaR}}^i_{\alpha}(\textbf{X})   = \E[g_i(Z) | \overline{F}_\textbf{X}(\textbf{X})= 1-\alpha]= \E[g_i(Z) | \overline{F}_{(Z,\ldots,Z)}(Z,\ldots,Z)= 1-\alpha]$. Since $\overline{F}_{(Z,\ldots,Z)}(u_1, \ldots, u_d)= \overline{F}_{Z}(\max_{i=1,\ldots, d} u_i),$ then
 $\overline{{\rm VaR}}^i_{\alpha}(\textbf{X})   =  \E[g_i(Z) | \overline{F}_{Z}(Z)= \alpha] =  {\rm VaR}_\alpha(X_i),$ for $i=1, \ldots, d$.    Hence the result.  $\Box$\vspace{0.17cm}

\begin{Remark}\label{independente VAR}
\noindent  For bivariate independent random couple $(X,Y)$,  Equations  \eqref{VAR XY integrale 1} and \eqref{VAR XY integrale 1 upper} become respectivley    \begin{eqnarray*}
\underline{\VAR}_{\alpha}^1(X,Y)   &=& \frac{1}{- \ln(\alpha)}\int_{\alpha}^1 \frac{\VAR_{\gamma}(X)}{\gamma} \, d\gamma, \\
\overline{\VAR}_{\alpha}^1(X,Y) &= &   \frac{1}{- \ln(1-\alpha)}\int_{0}^{1-\alpha} \frac{\VAR_{1-\gamma}(X)}{\gamma} \, d\gamma ,
\end{eqnarray*}
then,  obviously, in this case   the $X$-related component only depends  on the marginal behavior  of $X$.  For further details the reader is referred to Corollary 4.3.5 in  Nelsen (1999\nocite{NelsenLibro}). \end{Remark}

\subsection{Behavior of the multivariate VaR with respect to marginal distributions} \label{stochastic order copulas var}

\noindent
In this section we study the behavior of our  VaR  measures with respect to a change in  marginal distributions. Results presented below  provide natural multivariate  extensions  of some classical results in the univariate setting  (see, e.g.,  Denuit and Charpentier, 2004\nocite{charpentierbook}).

\begin{Proposition}\label{VAR invarinati con copula}
 Let  $\textbf{X}$ and $\textbf{Y}$  be  two $d$--dimensional  continuous  random vectors  satisfying  the  regularity conditions and with the same copula   $C$.     If $X_i  \buildrel d \over = Y_i $ then it holds that
 \begin{equation*}
\underline{{\rm VaR}}^i_{\alpha}(\textbf{X})= \underline{{\rm VaR}}^i_{\alpha}(\textbf{Y}),  \quad \mbox{ for all } \alpha \in (0,1),
\end{equation*}
and
 \begin{equation*}
 \overline{{\rm VaR}}^i_{\alpha}(\textbf{X})=  \overline{{\rm VaR}}^i_{\alpha}(\textbf{Y}),  \quad \mbox{ for all } \alpha \in (0,1).
\end{equation*}
\end{Proposition}
\vspace{0.1cm}

\noindent
The proof of the previous result directly comes down from   Equation   \eqref{VAR XY integrale 1} and  \eqref{VAR XY integrale 1 upper}.
From Proposition \ref{VAR invarinati con copula},  we remark that,  for a fixed copula $C$,  the $i$-th component $\underline{\VAR}_{\alpha}^i(\textbf{X})$ and $\overline{{\rm VaR}}^i_{\alpha}(\textbf{X})$ do not depend on marginal distributions of the other components $j$ with $j\neq i$. \vspace{0.2cm}

\noindent
In order to derive the next result, we  use the  definitions of stochastic orders  presented in Section~\ref{Notation}.
\begin{Proposition} \label{st order for VaR}
Let  $\textbf{X}$ and $\textbf{Y}$  be  two $d$--dimensional  continuous  random vectors  satisfying  the  regularity conditions  and with  the same copula   $C$.     If $X_i \preceq_{st} Y_i$  then it holds that
\begin{equation*}
\underline{{\rm VaR}}^i_{\alpha}(\textbf{X})\leq \underline{{\rm VaR}}^i_{\alpha}(\textbf{Y}), \quad \mbox{ for all } \alpha \in (0,1),
 \end{equation*}
 and
 \begin{equation*}
\overline{{\rm VaR}}^i_{\alpha}(\textbf{X})\leq \overline{{\rm VaR}}^i_{\alpha}(\textbf{Y}), \quad \mbox{ for all } \alpha \in (0,1).
 \end{equation*}
\end{Proposition}
\vspace{0.1cm}
\noindent
\textit{Proof: } The proof  comes down from  formulas \eqref{VAR XY integrale 1}- \eqref{VAR XY integrale 1 upper} and   Definition  \ref{def st order}.  Furtheremore, we remark that if $X_i \preceq_{st} Y_i$ then ${F}^{-1}_{X_i}(x) \leq{F}^{-1}_{Y_i}(x)$  for all $x$,  and $\overline{F}^{-1}_{X_i}(y) \leq \overline{F}^{-1}_{Y_i}(y)$ for all $y$. Hence the result.  $\Box$\vspace{0.17cm}

\noindent
Note that, the result in Proposition  \ref{st order for VaR} is consistent with the one-dimensional setting (see Section 3.3.1 in Denuit \emph{et al.}, 2005). Indeed, as in dimension  one, an increase of marginals with respect to the first order stochastic dominance yields an increase in the corresponding components of ${\rm VaR}_{\alpha}(\textbf{X})$.  \vspace{0.17cm}

As a result, in an economy with several interconnected financial institutions, capital required for one particular institution is affected by its own marginal risk.  But, for a fixed dependence structure, the solvency capital required for this specific institution does not depend on marginal risks bearing by the others. Then, our multivariate  VaR  implies a ``fair'' allocation of solvency capital with respect to individual risk-taking behavior. In other words, individual financial institutions may not have to pay more for risky business activities undertook by the others.

\subsection{Behavior  of multivariate  VaR  with respect to the dependence structure} \label{dependence structure copulas}

In this section we study the behavior of our VaR  measures with respect to a variation of the dependence structure, with unchanged marginal distributions.

\begin{Proposition} \label{st order for VaR avec U conditionel}
 Let  $\textbf{X}$ and $\textbf{X}^*$  be  two $d$--dimensional  continuous  random vectors  satisfying  the  regularity conditions and with the same margins $F_{X_i}$ and $F_{X^*_i}$, for $i=1, \ldots, d$,   and let $C$ {\rm(}resp. $C^*${\rm)} denote the copula function associated with $\textbf{X}$ {\rm(}resp. $\textbf{X}^*${\rm)} and $\overline{C}$ {\rm(}resp. $\overline{C}^*${\rm)}  the survival copula function associated with $\textbf{X}$ {\rm(}resp. $\textbf{X}^*${\rm)}.   \vspace{0.1cm}

Let $U_i=F_{X_i}(X_i)$,  $U_{i}^*=F_{{X_i}^*}(X_{i}^*)$,  $\textbf{U}= (U_1, \ldots, U_d)$ and $\textbf{U}^*= (U_1^*, \ldots, U_d^*)$.
 \begin{center}
If $\,\,[U_i | C(\textbf{U}) = \alpha] \preceq_{st} [U_{i}^* | C^*(\textbf{U}^*) = \alpha]\,\, \mbox{ then }\,\, \underline{{\rm VaR}}^i_{\alpha}(\textbf{X})\leq \underline{{\rm VaR}}^i_{\alpha}(\textbf{X}^*).$
\end{center}
\vspace{0.1cm}
Let $V_i=\overline{F}_{X_i}(X_i)$,  $V_{i}^*=\overline{F}_{{X_i}^*}(X_{i}^*)$,  $\textbf{V}= (V_1, \ldots, V_d)$ and $\textbf{V}^*= (V_1^*, \ldots, V_d^*)$.
\begin{center}
If $\,\,[V_i | \overline{C}(\textbf{V}) = 1-\alpha] \preceq_{st} [V_{i}^* | \overline{C}^*(\textbf{V}^*) = 1-\alpha]\,\, \mbox{ then }\,\, \overline{{\rm VaR}}^i_{\alpha}(\textbf{X}) \geq  \overline{{\rm VaR}}^i_{\alpha}(\textbf{X}^*).$
\end{center}
\end{Proposition}
\vspace{0.1cm}

\noindent
\textit{Proof: }  Let $U_1 \buildrel d \over = [U_i | C(\textbf{U}) = \alpha]$ and $U_2 \buildrel d \over = [U_{i}^{*} | C^*(\textbf{U}^*) = \alpha]$.  We recall that  $U_1 \preceq_{st} U_2$  if and only if $\E[f(U_1)]\leq \E[f(U_2)]$, for all non-decreasing function $f$ such that the expectations exist (see Denuit \emph{et al.}, 2005\nocite{Denuit}; Proposition 3.3.14).  We now choose   $f(u)=  F^{-1}_{X_i}(u)$,  for $u \in (0,1)$. Then, we obtain
\begin{center}
$\E[\, F^{-1}_{X_i}(U_i) | C(\textbf{U}) = \alpha \,] \leq \E[\, F^{-1}_{X_i}(U_{i}^{*}) | C^*(\textbf{U}^*) = \alpha \,]$,
\end{center}
\noindent
But  the right-hand side of the previous inequality is equal to $\E[\, F^{-1}_{X^{*}_{i}}(U_{i}^{*}) | C^*(\textbf{U}^*) = \alpha \,]$
since $X_i$ and $X_i^*$ have  the same  distribution.  Finally,  from  formula  \eqref{VAR XY integrale 1} we obtain   $\underline{{\rm VaR}}^i_{\alpha}(\textbf{X})\leq \underline{{\rm VaR}}^i_{\alpha}(\textbf{X}^*)$. \\
Let now $V_1 \buildrel d \over = [V_i | \overline{C}(\textbf{V}) = 1-\alpha]$ and $V_2 \buildrel d \over = [V_{i}^{*} | \overline{C}^*(\textbf{V}^*) = 1-\alpha]$.    We now choose  the non-decreasing function  $f(u)=  - \overline{F}^{-1}_{X_i}(u)$,  for $u \in (0,1)$. Since $X_i$ and $X_i^*$ have  the same  distribution, we obtain
\begin{center}
$\E[\, \overline{F}^{-1}_{X_i}(V_i) | \overline{C}(\textbf{V}) = 1-\alpha \,]  \geq  \E[\, \overline{F}^{-1}_{X_i}(V_i^{*}) | \overline{C}^*(\textbf{V}^*) = 1-\alpha \,]$,
\end{center}   Hence the result. $\Box$\\ \vspace{0.2cm}

\noindent
We now provide an illustration of  Proposition  \ref{st order for VaR avec U conditionel} in the case of  $d-$dimensional Archimedean copulas.

\begin{Corollary}\label{dependence_impact_Archi}
Consider a $d$--dimensional random vector  $\textbf{X}$,   satisfying  the  regularity conditions,  with  marginal distributions $F_{X_i}$,  for $i=1, \ldots, d$,  copula $C$ and survival copula $\overline{C}$.

 \begin{description}
   \item[] If $C$ belongs to one of the $d$-dimensional  family of Archimedean copulas introduced in Table \ref{kendall achimedeans}, an increase of the dependence parameter $\theta$ yields a decrease in each component of $\underline{{\rm VaR}}_{\alpha}(\textbf{X})$.
   \item[] If $\overline{C}$ belongs to one of the $d$-dimensional  family of Archimedean copulas introduced in Table \ref{kendall achimedeans}, an increase of the dependence parameter $\theta$ yields an increase  in each component of $\overline{{\rm VaR}}_{\alpha}(\textbf{X})$.
 \end{description}

\end{Corollary}

\textit{Proof: }
Let  $C_{\theta}$  and  $C_{\theta^*}$ be two Archimedean copulas of the same family with generator $\phi_{\theta}$ and $\phi_{\theta^*}$ such that $\theta\leq \theta^*$. Given Proposition \ref{st order for VaR avec U conditionel}, we have to check  that the relation $[U_i^* | C_{\theta^*}(\textbf{U}^*) = \alpha] \preceq_{st} [U_i | C_{\theta}(\textbf{U}) = \alpha]$ holds for all $i=1,\ldots, d$ where $(U_1, \ldots, U_d)$ and $(U_1^*, \ldots, U_d^*)$ are distributed (resp.) as $C_{\theta}$  and  $C_{\theta^*}$. However, using formula \eqref{Marginal_Cond_Distr_Archi}, we can readily prove that the previous relation can be restated as a decreasing condition on the ratio of generators $\phi_{\theta^*}$ and $\phi_{\theta}$, i.e.,
\begin{equation*}
[U_i^* | C_{\theta^*}(\textbf{U}^*) = \alpha] \preceq_{st} [U_i | C_{\theta}(\textbf{U}) = \alpha] \text{ for any }\alpha\in(0,1)   \iff \; \frac{\phi_{\theta^*}}{\phi_{\theta}} \text{ is a decreasing function.}
\end{equation*}
Eventually, we have check that, for all Archimedean family introduced in Table \ref{kendall achimedeans}, the function defined by  $\frac{\phi_{\theta^*}}{\phi_{\theta}}$ is indeed decreasing when $\theta \leq \theta^*$. We immediately obtain from Proposition \ref{st order for VaR avec U conditionel} that each component of   $\underline{{\rm VaR}}_{\alpha}(\textbf{X})$ is a decreasing function of $\theta$. The proof of the second statement of Corollary \ref{dependence_impact_Archi} follows trivially usign the same arguments. $\Box$\\

%We can prove that, for any  $d-$dimensional Archimedean copula whose Kendall distribution is given in Table \ref{kendall achimedeans}, an increase of the dependence parameter $\theta$ yields an increase in each component of $\overline{{\rm VaR}}_{\alpha}(\textbf{X})$. Indeed,
%Let  $\overline{C}_{\theta}$  and  $\overline{C}_{\theta^*}$ be two (survival) Archimedean copulas of the same family with generator $\phi_{\theta}$ and $\phi_{\theta^*}$ such that $\theta\leq \theta^*$.  From proof of Corollary \ref{dependence_impact_Archi} we have that
%$[U_i^* | \overline{C}_{\theta^*}(\textbf{U}^*) = 1-\alpha] \preceq_{st} [U_i | \overline{C}_{\theta}(\textbf{U}) = 1-\alpha] \text{ for any }\alpha\in(0,1)   \iff \; \frac{\phi_{\theta^*}}{\phi_{\theta}} \text{ is a decreasing function.} $
%And for all Archimedean family introduced in Table \ref{kendall achimedeans}, the function defined by  $\frac{\phi_{\theta^*}}{\phi_{\theta}}$ is  decreasing when $\theta \leq \theta^*$. We immediately obtain from Proposition \ref{st order for VaR avec U conditionel} that each component of   $\overline{{\rm VaR}}_{\alpha}(\textbf{X})$ is a increasing  function of $\theta$.

\begin{Example}
From Corollary \ref{dependence_impact_Archi}  the multivariate $\overline{\VAR}$ (resp. $\underline{\VAR}$) for copulas in Table  \ref{kendall achimedeans} is increasing  (resp. decreasing) with respect to the dependence parameter $\theta$ (coordinate by coordinate).  In particular, this means that, in the case of Archimedean copula, limit behaviors of dependence parameters  are associated with bounds for our multivariate risk measure.  For instance, let $(X,Y)$ be a bivariate random vector with a  Clayton dependence structure and fixed margins and $(\tilde{X},\tilde{Y})$ be a bivariate random vector with a Clayton survival copula and the same margins as $(X,Y)$. If we denote by  $\underline{{\rm VaR}}^1_{(\alpha, \theta)}(X, Y)$ (resp. $\overline{{\rm VaR}}^1_{(\alpha, \theta)}(\tilde{X},\tilde{Y})$)  the first component of  the lower-orthant VaR (resp.  upper-orthant VaR) when the dependence parameter is equal to $\theta$,  then the following comparison result holds for all ${\alpha \, \in \,(0,1)}$ and all $\theta \in (-1, \infty)$:
\begin{eqnarray*}
 && \overline{{\rm VaR}}^1_{(\alpha, -1)}(\tilde{X},\tilde{Y}) \leq  \overline{{\rm VaR}}^1_{(\alpha, \theta)}(\tilde{X},\tilde{Y}) \leq \overline{{\rm VaR}}^1_{(\alpha, +\infty)}(\tilde{X},\tilde{Y})\\
 &=& \underline{{\rm VaR}}^1_{(\alpha,  +\infty)}(X, Y)  \leq \underline{{\rm VaR}}^1_{(\alpha, \theta)}(X, Y) \leq \underline{{\rm VaR}}^1_{(\alpha, -1)}(X, Y). 
\end{eqnarray*}
Note that the upper bound corresponds to comonotonic random variables, so that $\overline{{\rm VaR}}^1_{(\alpha, + \infty)}(X, Y)$ $=$ $\underline{{\rm VaR}}^1_{(\alpha,  + \infty)}(X, Y)$ $=$ ${\rm VaR}_{\alpha}(X)= \alpha$, for a random vector ($X, Y$) with uniform marginal distributions.
\end{Example}

\subsection{Behavior of  multivariate  VaR   with respect to risk  level} \label{proprieta VAR PRD}

In order to study the behavior of the multivariate lower-orthant \emph{Value-at-Risk} with respect to risk  level  $\alpha$, we need to introduce the \emph{positive regression dependence} concept.  For a bivariate random vector $(X,Y)$ we mean by positive dependence that $X$ and $Y$ are likely to be large or to be
small together. An excellent presentation of  positive dependence concepts can be found in Chapter 2 of the book by Joe (1997)\nocite{Joe}.  The  positive dependence concept  that will be used in the sequel has been called  \emph{positive regression dependence} (PRD) by Lehmann (1966)\nocite{Lehmann}  but most of the authors use the term \emph{stochastically increasing} (SI) (see Nelsen, 1999\nocite{NelsenLibro}; Section 5.2.3).

\begin{Definition}[Positive regression dependence]\label{PRD}
\begin{sloppypar}
A bivariate random vector ${(X,Y)}$ is said to admit positive regression dependence with respect to $X$, {\rm PRD(}$Y | X${\rm)}, if
 \begin{equation}\label{PRD formula}
[Y | X=x_1] \preceq_{st} [Y | X=x_2],\quad \forall \,x_1 \leq x_2.
  \end{equation}\end{sloppypar}
\end{Definition}
\vspace{0.15cm}

\noindent
Clearly condition in  \eqref{PRD formula} is a positive dependence notion (see Section 2.1.2 in Joe, 1997\nocite{Joe}).
%In the following, a few examples are used to illustrate the dependence concept  in Definition  \ref{PRD}.
%\begin{Example}
%  \item  Let   $f(x_1,x_2)= \frac{1}{(2 \pi)(1-\rho^2)^{1/2}} \, \rme^{\frac{-\frac{1}{2}(x_{1}^2+x_{2}^2- 2 \rho \, x_1 \,x_2)}{(1- \rho^2)}}$ be the bivariate  normal density  with $\rho \in (-1,1)$. Then if $\rho >0$ we obtain {\rm PRD(}$X_2 | X_1${\rm)} (see Example 2.1 in Joe, 1997\nocite{Joe}).
%  \item   We consider the bivariate  copula $(4.2.6)$ in  Table 4.1 of Nelsen (1999)\nocite{NelsenLibro}: $C(u,v)=1-[(1-u)^\theta + (1-v)^\theta - (1-u)^\theta \,  (1-v)^\theta]^{\frac{1}{\theta}}$, with $\theta \in [1, \infty)$.  Then   {\rm PRD(}$V | U${\rm)},  for each $\theta \in [1, \infty)$ (see Example 2.3 in Joe, 1997\nocite{Joe}).
%  \end{Example}
%  \ares{Est-ce que les exemples precendents sont vraiment utiles?}

\noindent
From Definition \ref{PRD},  it is straightforward to derive the following result.
\begin{Proposition}\label{PRD Prop}
Consider a $d$--dimensional random vector  $\textbf{X}$,   satisfying  the  regularity conditions,  with  marginal distributions $F_{X_i}$,  for $i=1, \ldots, d$,  copula $C$ and survival copula $\overline{C}$.  Let $U_i=F_{X_i}(X_i)$, $\textbf{U}= (U_1, \ldots, U_d)$, $V_i= \overline F_{X_i}(X_i)$ and $\textbf{V}= (V_1, \ldots, V_d)$. Then it holds that :\vspace{0.1cm}
 \begin{itemize}
  \item[] If $(U_i,C(\textbf{U}))$ is $\mbox{PRD}(U_i| C(\textbf{U}))$ then $\underline{{\rm VaR}}_{\alpha}^i(\textbf{X})$ is a   non-decreasing function of $\alpha$. \\
  \item[] If $(V_i, \overline{C}(\textbf{V}))$ is $\mbox{PRD}(V_i |  \overline{C}(\textbf{V}))$ then $\overline{{\rm VaR}}_{\alpha}^i(\textbf{X})$ is a  non-decreasing function of $\alpha$.\vspace{0.15cm}
\end{itemize}
 \end{Proposition}

\noindent
\textit{Proof: } If $\alpha_1\leq \alpha_2$, we have
$[U_i | C(\textbf{U})=\alpha_1] \preceq_{st} [U_i | C(\textbf{U})=\alpha_2]$ and $[V_i | \overline{C}(\textbf{V})=1-\alpha_2] \preceq_{st} [V_i | \overline{C}(\textbf{V})=1-\alpha_1]$.  As in the proof of Proposition \ref{st order for VaR avec U conditionel},
 \begin{center}
$\E[\, F^{-1}_{X_i}(U_i) | C(\textbf{U}) = \alpha_1\,] \leq \E[\, F^{-1}_{X_i}(U_i)  | C(\textbf{U}) = \alpha_2 \,]$.
\end{center}
and
 \begin{center}
$\E[\, \overline{F}^{-1}_{X_i}(V_i) | \overline{C}(\textbf{V}) = 1-\alpha_1\,] \leq \E[\, \overline{F}^{-1}_{X_i}(V_i)  | \overline{C}(\textbf{V}) = 1-\alpha_2 \,]$.
\end{center}

 Then $\underline{\VAR}_{\alpha_1}^i(\textbf{X})\leq \underline{\VAR}_{\alpha_2}^i(\textbf{X})$ and  $\overline{\VAR}_{\alpha_1}^i(\textbf{X})\leq \overline{\VAR}_{\alpha_2}^i(\textbf{X})$,   for any $\alpha_1 \leq \alpha_2$ which proves that $\underline{\VAR}_{\alpha}^i(\textbf{X})$ and $\overline{\VAR}_{\alpha}^i(\textbf{X})$  are  non-decreasing  functions of $\alpha$.  $\Box$\vspace{0.17cm}

\noindent
Note that behavior of the  multivariate  \underline{VaR} with respect to a change in the risk level does not depend on marginal distributions of $\textbf{X}$.\vspace{0.17cm}

The following result proves  that assumptions of Proposition \ref{PRD Prop}  are satisfied in the large  class of  $d$-dimensional   Archimedean  copulas.

\begin{Corollary}\label{PRD Prop Archimediennes}
Consider a $d$--dimensional random vector  $\textbf{X}$,   satisfying  the  regularity conditions,  with  marginal distributions $F_{X_i}$,  for $i=1, \ldots, d$,  copula $C$ and survival copula $\overline{C}$.
\begin{center} If $C$ is a $d$-dimensional  Archimedean  copula, then  $\underline{{\rm VaR}}_{\alpha}^i(\textbf{X})$ is a   non-decreasing function of $\alpha$. 
\end{center}
\begin{center} If $\overline{C}$ is a $d$-dimensional  Archimedean  copula, then  $\overline{{\rm VaR}}_{\alpha}^i(\textbf{X})$ is a   non-decreasing function of $\alpha$.
\end{center}
\end{Corollary}
\noindent
\textit{Proof: } Let $U_i=F_{X_i}(X_i)$, $\textbf{U}= (U_1, \ldots, U_d)$, $V_i= \overline F_{X_i}(X_i)$ and $\textbf{V}= (V_1, \ldots, V_d)$. If $C$ is the copula of $X$, then $\textbf{U}$ is distributed as $C$ and if $C$ is Archimedean, $\Pbb[U_i >  u\, | \, C(\textbf{U}) = \alpha]$   is a non-decreasing  function of $\alpha$ from formula  \eqref{Marginal_Cond_Distr_Archi}. In addition, if $\overline{C}$ is the survival copula of $X$, then $\textbf{V}$ is distributed as $\overline{C}$ and if $\overline{C}$ is Archimedean, $\Pbb[V_i >  u\, | \, \overline{C}(\textbf{V}) = \alpha]$  is a non-decreasing  function of $\alpha$ from the same argument. The result then derives from  Proposition \ref{PRD Prop}. $\Box$\vspace{0.17cm}

\section*{Conclusion and perspectives}\label{Conclusions}
\begin{sloppypar}
\noindent
%\ares{Conclusions a recrire  completement}
In this paper, we proposed two multivariate extensions of the classical Value-at-Risk for continuous random vectors.  As in the Embrechts and Puccetti (2006)'s approach, the introduced risk measures are based on multivariate generalization of quantiles but they are able to quantify risks in a much more parsimonious and synthetic way:  the risk of a $d$-dimensional portfolio is evaluated by a point in $\Real^d_+$. The proposed multivariate risk measures may be useful for some applications where risks are heterogeneous in nature or because they cannot be diversify away by an aggregation procedure.\\

\noindent
We  analyzed  our multivariate risk measures in several directions. Interestingly, we showed  that many properties satisfied by the univariate VaR expand to the  two proposed multivariate versions under some conditions. In particular, the \emph{lower-orthant} VaR   and the \emph{upper-orthant} VaR   both satisfy  the positive homogeneity and the translation invariance property which are parts of the classical axiomatic  properties of Artzner \emph{et al.}  (1999).  Using  the theory of stochastic ordering, we also analyzed the effect of some risk perturbations on these measures. In the same vein as for the univariate VaR, we proved that an increase of marginal risks yield an increase of the multivariate VaR.  We also gave the condition under which an increase of the risk level tends to increase components of the proposed multivariate extensions and we show that these conditions are satisfied for $d$-dimensional Archimedean copulas. We also study the effect of dependence between risks on individual contribution of the multivariate VaR and we prove that for different   families of Archimedean copulas, an increase of the dependence parameter tends to lower the components of the \emph{lower-orthant}  VaR whereas it widens the components of the \emph{upper-orthant} VaR. At the extreme case where risks are perfectly dependent or comonotonic, our multivariate risk measures are equal to the vector composed of univariate risk measures associated with each component.\\
%This feature is in line with the observation made by Zhou (2010)\nocite{Zhou}:
%\emph{``When regulating a system consisting of similar institutions, or in other words, the system is highly interconnected, considering a micro-prudential regulation can be sufficient for reducing the overall systemic risk."}   (Zhou, 2010\nocite{Zhou}).\vspace{0.17cm}

\noindent Due to the fact that the Kendall distribution is not known analytically for elliptical random vectors, it is still an open question whether components of our proposed measures are increasing with respect to the risk level for such dependence structures. However, numerical experiments in the case of Gaussian copulas support this hypothesis. More generally,  the extension of the  McNeil and Ne\v{s}lehov\'{a}'s representation  (see Proposition \ref{McNeil_Neslehova})  for a generic copula $C$ and the study of the behavior of distribution $[U | C(\textbf{U})=\alpha]$, with respect to $\alpha$,  are potential improvements to this paper that will be investigated in a future work.\\

\noindent In a future perspective, it should also be interesting to discuss the extensions of our risk measures to the case of discrete distribution functions, using ``discrete level sets'' as multivariate definitions of quantiles. For further details the reader is referred, for instance, to  Laurent (2003\nocite{Laurent}).  Another subject of future research should be to introduce a similar multivariate extension but for \emph{Conditional-Tail-Expectation} and compare the proposed VaR and CTE measures with existing multivariate generalizations of risk measures,  both theoretically and experimentally.  An article is in preparation in this sense. \end{sloppypar}

\vspace{0.3cm}

\begin{sloppypar}
\noindent
 \textbf{Acknowledgements:}
    The authors thank an  anonymous referee for   constructive  remarks and valuable suggestions to improve the paper. The authors thank V\'{e}ronique Maume-Deschamps and  Cl\'{e}mentine Prieur  for their comments and help and Didier Rulli{\`e}re for fruitful discussions. This work has been partially supported by the French research national agency (ANR) under the reference ANR-08BLAN-0314-01. Part of this work also benefit from the support of the MIRACCLE-GICC project.
\end{sloppypar}

\vspace{0.3cm}

\textbf{References:}
%\scriptsize
\bibliographystyle{abbrv}
\addcontentsline{toc}{section}{Bibliography}
\bibliography{biblio}

\end{document}